\documentclass[11pt]{article}
\pdfoutput=1
\usepackage{jheppub}
\usepackage{graphicx}
\usepackage{amssymb}
\usepackage{bbm}
\usepackage{microtype}
\usepackage{mathrsfs}
\usepackage{amsmath,amssymb}
\usepackage{slashed}
\usepackage{hyperref}
\usepackage{circuitikz}
\usepackage{caption}
\usepackage{xcolor}
\usepackage{dsfont}
\usepackage{verbatim}
\usepackage{subfig}
\usepackage{amsmath,amssymb,amscd,amsfonts,mathtools,csquotes}
\usepackage{xifthen}
\usepackage{xcolor}
\definecolor{markgreen}{RGB}{230,243,230}
\definecolor{darkolivegreen}{rgb}{0.33, 0.42, 0.18}
\definecolor{darkpastelgreen}{rgb}{0.01, 0.75, 0.24}
\DeclareMathOperator{\Tr}{Tr}

\usepackage[toc,page]{appendix}
\usepackage{epsfig}
\usepackage{epstopdf}
\usepackage{latexsym}
\usepackage{graphicx}
\usepackage{booktabs}
\usepackage{bbm}
\usepackage{color}
\usepackage{physics}
\usepackage{tensor}
\usepackage{verbatim}
\usepackage{subfig}
\usepackage{tikz}
\usepackage{ifthen}
\usepackage{array}
\usetikzlibrary{matrix}
\usetikzlibrary{decorations.markings,calc,shapes,decorations.pathmorphing,patterns,decorations.pathreplacing,arrows,arrows.meta}
\usetikzlibrary{positioning}
\usepackage{hyperref}
\usetikzlibrary{patterns}
\usetikzlibrary{positioning}
\newdimen\mydim
\pdfoutput=1
\makeatletter
\def\@fpheader{\relax}

\newcommand*{\ov}[1]{
  $\m@th\overline{\mbox{#1}}$
}
\newcommand*{\ovA}[1]{
  $\m@th\overline{\mbox{#1}\raisebox{3mm}{}}$
}
\newcommand*{\ovB}[1]{
  $\m@th\overline{\mbox{#1\rule{0pt}{3mm}}}$
}
\newcommand*{\ovC}[1]{
  $\m@th\overline{\mbox{#1\strut}}$
}
\newcommand*{\ovD}[1]{
  $\m@th\overline{\mbox{#1\vphantom{\"A}}}$
}
\newcommand*{\ovE}[1]{
  $\m@th\overline{\raisebox{0pt}[1.2\height]{#1}}$
}
\newcommand*{\ovF}[1]{
  $\m@th\overline{\raisebox{0pt}[\dimexpr\height+1mm\relax]{#1}}$
   Package `calc' can be used as alternative for `\dimexpr'.
}
\newcommand*{\ovG}[1]{
  $\m@th\overline{\raisebox{0pt}[\dimexpr\height+1mm\relax]{#1\vphantom{A}}}$
}
\makeatother
\newboolean{showcomments}
\setboolean{showcomments}{false}
\newcommand\rem[1]{\ifthenelse{\boolean{showcomments}}{{#1}}{}}
\newcommand{\be}{\begin{equation}}
\newcommand{\ee}{\end{equation}}

\usepackage{xifthen}
\newcommand{\dalembert}[1][]{\ifthenelse{\isempty{#1}}{\Box}{#1\Box}}
\usetikzlibrary{decorations.pathmorphing}
\tikzset{snake it/.style={decorate, decoration=snake}}

\title{\Large Making the Case for Massive Islands}
\author{Hao Geng}
\affiliation{Gravity, Spacetime, and Particle Physics (GRASP) Initiative, Harvard University, 17 Oxford St., Cambridge, MA, 02138, USA.}

\emailAdd{haogeng@fas.harvard.edu}

\abstract{It has been established in both high and low spacetime dimensions that existing constructions of entanglement islands, at least in the study of black holes, are all in the context of massive gravity. In fact, later studies realized that the graviton mass is not an accident and it is necessary for the consistency of the holographic interpretation for entanglement islands. An important lesson we learned from these studies is that the graviton mass is a manifestation of the deep relationship between quantum entanglement and emergent geometry (EPR=ER). Nevertheless, various interesting questions and counter-arguments exist regarding whether this connection is of any relevance, either by modifying the definition of entanglement islands or denying the graviton mass as a physically relevant concept. In this paper, we will revisit the necessity of graviton mass for the consistency of entanglement islands and address the above questions in and out of the literature. Since the calculations in higher dimensions are already well understood, we will start with a demonstration of massive islands in a frequently used lower dimensional model-- AdS$_{2}$ Jackiw-Teitelboim (JT) gravity coupled with transparent conformal matter fields, for which the relationship between energy dissipation and the graviton mass manifests. We will then discuss the necessity of graviton mass for entanglement islands in general situations and clarify the physical meaning of graviton mass. In the end, we will comment on the attempts to construct entanglement islands in massless gravity and suggest the search for a new type of entanglement wedges in holography.

The purpose of this paper is to summarize the recent progress, clarifying potential confusions and convey key messages to both experts and non-experts in the field.}

\begin{document}
\maketitle
\flushbottom
\pagebreak

\section{Introduction}
The original Hawking's black hole information paradox \cite{Hawking:1974sw,Hawking:1976ra} states that for an evaporating black hole formed from collapse the fine-grained entropy of the system is not conserved and is hence a violation of quantum unitarity. This is because the fine-grained entropy of the initial matter configuration can be tuned arbitrarily low, but at the end of the evaporation one naively ends up with a chunk of thermal radiation whose entropy is purely thermal and is therefore large. This paradox is later refined by Page \cite{Page:1993wv} with a speculative resolution as the so-called \textit{Page curve}. 

The idea of Page is that in the above process of black hole formation and evaporation the apparent breakdown of quantum unitarity appears at a particular instant which is called the \textit{Page time}. Page interpreted the entropy of the black hole radiation as the entanglement entropy between the emitted black hole radiation and the remanent black hole, and further noticed that the apparently growing entropy of the black hole radiation will eventually bypass the decreasing Bekenstein-Hawking entropy of the evaporating remanent black hole. The instant this bypassing happens is the moment the quantum unitarity breaks down, i.e. the Page time, as the Bekenstein-Hawking entropy is upper bounding the dimension of the black hole Hilbert space and it cannot be smaller than the entanglement entropy between the black hole and any system entangled with it. Hence, Page suggested that the resolution of the information paradox relies on showing that the entropy of the black hole radiation in fact follows the Page curve, i.e. after the Page time this entropy should go down rather than staying as going up. This question had then become the stumbling block of theoretical physics as the fine-grained entropy is a highly nonlinear quantity and it was not so clear how to carefully calculate it with a result following the Page curve.

Further confusion kicked in after the proposal of the \textit{holographic principle} \cite{tHooft:1993dmi,Susskind:1994vu} and its explicit realization in string theory-- the \textit{AdS/CFT correspondence} \cite{Maldacena:1997re,Gubser:1998bc,Witten:1998qj}. The AdS/CFT correspondence states that quantum gravity theory in asymptotically anti-de Sitter (AdS) space is dual to a conformal field theory (CFT) living on its asymptotic boundary. Hence, the quantum gravity theory is fully unitary. This suggests that if one considers the black hole formation and evaporating process, as Hawking's original information paradox, in the AdS space, the entropy of the system should be a constant in time due to unitarity. Thus, the information paradox is resolved as a direct implication of the holographic principle. Nevertheless, to study the above problem suggested by Page, one has to separate the black hole radiation from the remanent black hole, which though was not so clear how to achieve in the AdS/CFT correspondence. This is because the CFT encodes the full system inside the AdS and it is not so clear which degrees of freedom in CFT should be identified as the radiation and which should be as the black hole. In fact, such a separation is rather unnatural from the AdS point of view if one carefully considers gravitational effects \cite{Laddha:2020kvp,Raju:2020smc,Geng:2021hlu,Raju:2021lwh}. 

Important progress was made in \cite{Almheiri:2019psf,Penington:2019npb} by realizing that the problem suggested by Page can be studied in the AdS/CFT correspondence by coupling the gravitational AdS spacetime with a nongravitational bath on its asymptotic boundary. This setup was originally proposed by Karch and Randall in \cite{Karch:2000ct,Karch:2000gx} for a different purpose. In this setup, one can think of the bath as absorbing the radiation from the black hole and so the black hole and its radiation are now nicely separated. This is because, applying the AdS/CFT correspondence, this setup is dual to a setup with a CFT coupled to a bath residing on the boundary of the bath. Thus, one can think of the black hole as the CFT and the emitted radiation as the bath. As a result, the properties of the black hole radiation can be easily understood by studying the nongravitational bath. For example, one can take a subregion $R$ of the bath away from the place where the bath and the AdS are coupled, i.e. away from the CFT, as the emitted Hawking radiation at early times and compute its entanglement entropy as a function of time.

From the point of view of CFT coupled with a bath, this entanglement entropy will certainly obey a unitary Page curve. Nevertheless, from the perspective of the gravitational AdS coupled with the bath, it is not so straightforward why this entropy will follow the unitary Page curve as opposed to Hawking's original result.
Interestingly, a careful application of the replica trick with gravitational effects properly incorporated implies that the entanglement entropy of $R$ can be computed using the \textit{quantum extremal surface} formula \cite{Engelhardt:2014gca}
\begin{equation}
    S(R)=\min_{ \mathcal{I}}\Big[S_{\text{QFT}}(R\cup\mathcal{I})+\frac{A(\partial\mathcal{I})}{4G_{N}}\Big]\,,\label{eq:islandformula}
\end{equation}
where $S_{\text{QFT}}$ denotes the subregion entanglement entropy in a fixed nongravitational background and $\mathcal{I}$ is a closed subregion inside the gravitational AdS, along the same Cauchy slice as $R$, and it is called the \textit{entanglement island} of $R$. The gravitational effects leading to this formula are the so-called \textit{replica wormholes} \cite{Almheiri:2019qdq,Penington:2019kki,Geng:2024xpj,Geng:2025efs}. The formula produces the unitary Page curve as one can think of $\mathcal{I}$ as the black hole interior and the entanglement between $R$ and $\mathcal{I}$ will accumulate with time such that $S_{\text{QFT}}(R)$ by itself is able to compete with $\frac{A(\partial\mathcal{I})}{4G_{N}}$ at late times, and eventually dominates over it, so the result of the minimization in Equ.~(\ref{eq:islandformula}) produces a nontrivial $\mathcal{I}$ and the resulting $S(R)$ is no longer increasing as $S_{\text{QFT}}(R)$. Thus, the moment a nontrivial entanglement island $\mathcal{I}$ emerges is precisely the Page time. 

Besides reproducing a unitary Page curve, the QES formula has a rather deep implication. This is because it further states that the information inside the island $\mathcal{I}$ is fully encoded in $R$. In the context of black holes, this means that the information in the black hole interior, which island $\mathcal{I}$ largely overlaps with, is fully encoded in its early-time radiation. This is interesting for two reasons: 1) this resolves the firewall paradox \cite{Mathur:2009hf,Mathur:2012jk,Mathur:2012np,Almheiri:2012rt,Almheiri:2013hfa}; 2) this suggests a nonlocal information encoding scheme as $\mathcal{I}$ and $R$ are spacelike separated from each other and they are living in different universes. We are mostly interested in the second point. This is because a naive application of the AdS/CFT correspondence would imply that all the AdS spacetime including $\mathcal{I}$ is encoded in the dual CFT.\footnote{This point was emphasized by Juan Maldacena in a discussion.} But the QES formula implies that the physics inside $\mathcal{I}$ is in fact factorized from the CFT as it is encoded in another nongravitational system $R$ which is factorized from the CFT. This sounds rather astonishing, as if the gravitational theory inside the AdS is not significantly modified due to the bath coupling, the AdS/CFT correspondence should apply and so all the physics inside AdS including $\mathcal{I}$ should indeed be encoded in the CFT. Thus, this observation already suggests that something interesting happens to the gravitational theory due to the bath coupling.

If one reminds himself/herself of the original work from Karch and Randall \cite{Karch:2000ct,Karch:2000gx},\footnote{Also the work from Porrati \cite{Porrati:2003sa} and its later refinement and generalizations \cite{Duff:2004wh,Aharony:2006hz}.} then one realizes that the graviton inside AdS becomes massive due to the bath coupling. Thus, one naturally conjectures that the very existence of entanglement islands would have a connection with the nonvanishing graviton mass \cite{Geng:2020qvw}. This is the \textit{massive islands conjecture} proposed by Geng and Karch.\footnote{This is also speculated by Donald Marolf in some private discussions. We thank Daniel Jafferis for pointing this out.} Further evidence for this conjecture was found in \cite{Geng:2020fxl} which noticed that there is no entanglement island in a controllable model of massless gravity due to diffeomorphism invariance. Motivated by this work, \cite{Geng:2021hlu} then precisely formulated the tension between entanglement islands and standard massless graviton using the diffeomorphism constraints. The diffeomorphism constraints are quantum versions of the classical gravitational Gauss' law. One can already see the tension between entanglement islands and standard massless gravity using the gravitational Gauss' law. This is because the gravitational Gauss' law states that one can detect an energetic excitation inside a gravitational universe by measuring gravitational potential near its boundary due to the long-range nature of gravity. Thus, in the island scenario if the gravitational theory is the standard massless gravity then one can detect excitations inside the island from its complement inside the AdS. This directly contradicts the factorization between the island $\mathcal{I}$ and the CFT. However, when the graviton is massive, the gravitational force becomes short-ranged due to the Yukawa type exponential decay, and the above contradiction is automatically avoided.

   Recent progress in \cite{Geng:2025rov} further shows that the nonzero graviton mass in the island scenario is also crucial for another consistency condition. This consistency condition is based on the observation that energetic excitations inside the island should be detectable by the bath Hamiltonian. Thus, operators inside the island should have nonzero commutators with the bath Hamiltonian. Nevertheless, this condition is not so obviously satisfied as the island and the bath are spacelike separated from each other with a buffer zone in between. Thus, naively if one ignores gravitational effects, operators inside the island should commute with operators in the bath including the bath Hamiltonian. As we will explain in later discussions in the main text, this nonzero commutator is achieved due to the nonzero graviton mass and it has a deep implication to the connection between quantum entanglement and emergent spacetime geometry (EPR=ER).\footnote{This explanation is provided in Sec.~\ref{sec:review}. However, we will refer interested readers to \cite{Geng:2025rov} for technical details.}

However, interesting questions and possible counter-arguments about the connection between graviton mass and entanglement islands exist in the community. Some of these arguments are recently collected in \cite{Antonini:2025sur} with possible counter-examples to massive islands. The aim of the current paper is to revisit the massive island story with a systematic summary and analysis of these questions/arguments and possible counter-examples. The hope is to illuminate the recent progress in quantum gravity to a broader community of physicists. Though, we refer experts directly to Sec.~\ref{sec:wormholemass}, Sec.~\ref{sec:robust}, Sec.~\ref{sec:meaningmass} and Sec.~\ref{sec:attempts} for the analysis of various questions and counter-arguments about massive islands.

We should mention that in the discussion so far in the introduction, we consider entanglement islands in the context of evaporating black holes to motivate their connection with the black hole information paradox. In fact, entanglement islands ubiquitously exist in the setup with gravitational AdS coupled with a nongravitational bath. Following \cite{Geng:2025rov}, we will call this setup the \textit{island model}. To understand the key physics of entanglement islands, it is much simpler to consider the case with an empty AdS coupled with a bath. In this case, explicit calculations can be done and the results are easily generalized to other cases.

This paper is organized as follows: 
\begin{enumerate}
    \item In Sec.~\ref{sec:revisit} we will revisit the fact that graviton is massive in the island scenario due to the bath coupling. The emphasis is on the fact that the bath coupling spontaneously breaks the diffeomorphism symmetry in the gravitational AdS and thus the graviton picks up a mass as a result of the Higgs mechanism. Since the calculation has been well-understood in higher dimensions, we will use lower-dimensional models to demonstrate this Higgs mechanism. The result resolves a speculation identifying the graviton mass with the characteristic energy scale associated with energy dissipation into the bath.\footnote{As will be discussed later, this point was firstly brought up to us by Henry W. Lin \cite{Almheiri:2021up}.}
    \item  In Sec.~\ref{sec:in general} we will revisit the general reason why the graviton mass is both essential and necessary for entanglement islands. We will see that this connection is in fact a manifestation of the deep relationship between quantum entanglement and emergent geometry (EPR=ER).
    \item In Sec.~\ref{sec:meaningmass} we will articulate the physical meaning of the graviton mass addressing various concerns regarding if the graviton mass is locally measurable inside the bulk of AdS. In Sec.~\ref{sec:attempts} we will analyze various attempts and possible examples in constructing entanglement islands in massless gravity theories.
    \item  We conclude this paper with discussions in Sec.~\ref{sec:conclusion}.
\end{enumerate}

\section{Massive Islands in the CFT Coupled JT Model}\label{sec:revisit}
In this section, we revisit the essential physics of massive islands using a simple lower-dimensional model. This model is Jackiw-Teitelboim (JT) gravity in AdS$_{2}$ coupled with conformal matter fields. The conformal matter fields are described by a conformal field theory with central charge $c$ and it is coupled to the same conformal field theory on a half Minkowski space. This half Minkowski space is not gravitational and it is glued to the AdS$_{2}$ along its asymptotic boundary.\footnote{In fact, this gluing is subtle for the JT model as the AdS$_{2}$ asymptotic boundary is cutoff. The dynamics of JT gravity is reflected in the fluctuation of the shape of this cutoff boundary. So the gluing map is also dynamical \cite{Almheiri:2019qdq}. This subtlety makes the application of this particular model to the study of island a bit unnatural but it is not relevant to the discussion of this paper and we will ignore it hereafter.} The energy-momentum flux of the conformal field theory freely flows between the AdS$_{2}$ and the half Minkowski space. Thus, the half Minkowski space can be thought of as a bath. This model is frequently used in the study of entanglement islands in the literature.

As opposed to the higher-dimensional setups, the meaning of graviton mass is not apparent in this lower-dimensional model, as there is no dynamical graviton in 2d. However, as it has been understood in \cite{Geng:2023ynk,Geng:2023zhq,Geng:2025rov}, the essence of the massive islands story is really that the bath coupling induces a spontaneous breaking of the AdS diffeomorphism symmetries which generates a St\"{u}ckelberg mass term for the metric fluctuation $h_{\mu\nu}$. This is a quantum effect and the St\"{u}ckelberg mass term modifies the diffeomorphism constraints in the AdS which allows operators to be consistently localized inside the islands. We will discuss details on the operators inside islands in Sec.~\ref{sec:necc}. For the purpose of this section, we will try to obtain this quantum induced St\"{u}ckelberg mass term for the metric fluctuation $h_{\mu\nu}$ in AdS$_{2}$.

\subsection{Direct Calculation of the Mass}\label{sec:direct}
The system we are interested is described by the action  $S_{\text{tot}}[\phi,\psi,g]=S_{0}+S[\phi;g]+S_{\text{matter}}[\psi;g]$ in the AdS$_{2}$ part with
\begin{equation}
    \begin{split}
S_{0}&=\frac{\phi_{0}}{16\pi G_{N}}\Big[\int_{\mathcal{M}}d^{2}x\sqrt{-g}R+2\int_{\partial\mathcal{M}}K\Big]\,,\\S[\phi;g]&=\frac{1}{16\pi G_{N}}\Big[\int_{\mathcal{M}}d^{2}x\sqrt{-g}\phi(R+2)+2\int_{\partial\mathcal{M}}\phi_{\partial}K\Big]\,,\\S_{\text{matter}}[\psi;g]&=S_{\text{CFT}}[g]\,,\label{eq:action}
    \end{split}
\end{equation}
where $\mathcal{M}$ denotes the AdS$_{2}$ manifold and $\partial\mathcal{M}$ denotes its boundary. We used $\psi$ to collectively denote the CFT fields, $\phi$ is called the dilaton field and $S_{0}$ is purely topological. We have set the AdS$_2$ length scale $l_{AdS}$ to one. The path integral of this system is given by
\begin{equation}
    Z=\int D[g]D[\phi]D[\psi]e^{iS_{\text{tot}}[\phi,\psi,g]}\,,
\end{equation}
from which we can see that the dilaton $\phi$ is a Lagrange multiplier which imposes the constraint that
\begin{equation}
    R=-2\,.
\end{equation}
Thus, the spacetime geometry is fixed to be hyperbolic. We are only interested in gravitational perturbation theory around the simplest topology. Hence, we can take the spacetime geometry to be the AdS$_{2}$ Poincar\'{e} patch
\begin{equation}
    ds^2=g^{0}_{\mu\nu}dx^{\mu}dx^{\nu}=\frac{-dt^2+dx^2}{x^2}=-\frac{4dx^{+}dx^{-}}{(x^{+}-x^{-})^2}\,,\quad x^{\pm}=t\pm x\,,\label{eq:poincare}
\end{equation}
where $x\in[\epsilon,\infty)$ and $\epsilon\rightarrow0$. Furthermore, the metric is perturbatively expanded around the background $g^{0}_{\mu\nu}$ to the leading nontrivial order as
\begin{equation}
    g_{\mu\nu}=g^{0}_{\mu\nu}+\sqrt{16\pi G_{N}}h_{\mu\nu}\,.
\end{equation}

For the purpose of extracting the mass, we only have to focus on the interaction between the CFT and the perturbative metric fluctuation $h_{\mu\nu}$. Thus, we can focus on the following path integral
\begin{equation}
    Z_{h}=\int D[h]D[\psi]e^{iS_{\text{matter}}[\psi;g^{0}]+iS_{\text{int}}[\psi,h;g^{0}]}\,,\label{eq:Zh}
\end{equation}
for which the interaction term is
\begin{equation}
    S_{\text{int}}[\psi,h;g^{0}]=\frac{\sqrt{16\pi G_{N}}}{2}\int d^2x\sqrt{-g^{0}}h_{\mu\nu} T^{\mu\nu}_{\text{CFT}}\,,
\end{equation}
where we used the convention that $T^{\mu\nu}_{\text{CFT}}=\frac{2}{\sqrt{-g}}\frac{\delta S_{\text{matter}}[\psi;g]}{\delta g_{\mu\nu}}\big\vert_{g_{\mu\nu}=g^{0}_{\mu\nu}}$. We note that to the leading nontrivial order in $G_{N}$, we have the diffeomorphism transform of the relevant fields and the background
\begin{equation}
\begin{split}
    h'_{\mu\nu}(x)&= \frac{\partial x^{'\rho}}{\partial x^{\mu}}\frac{\partial x^{'\sigma}}{\partial x^{\nu}}h_{\rho\sigma}(x')+\nabla_{\mu}\epsilon_{\nu}(x)+\nabla_{\nu}\epsilon_{\mu}(x)\,,\\
\psi'(x)&=\psi(x')\,,\quad g^{'0}_{\mu\nu}(x)=\frac{\partial x^{'\rho}}{\partial x^{\mu}}\frac{\partial x^{'\sigma}}{\partial x^{\nu}}g^{0}_{\rho\sigma}(x')\,,
    \end{split}
\end{equation}
where $x^{'\mu}=x^{\mu}+\sqrt{16\pi G_{N}}\epsilon^{\mu}(x)$ and $\nabla_{\mu}$ is the torsion free and metric compatible covariant derivative with respect to $g^{0}_{\mu\nu}$. In fact, the above transform is equivalent to only transform $h_{\mu\nu}(x)$ as
\begin{equation}
    h_{\mu\nu}(x)\rightarrow h_{\mu\nu}(x)+\nabla_{\mu}\epsilon_{\nu}(x)+\nabla_{\nu}\epsilon_{\mu}(x)\,,\label{eq:nontrivialTr}
\end{equation}
with other fields not transformed. In the following, we will focus on the transform Equ.~(\ref{eq:nontrivialTr}). 

It is important to notice that the path integral Equ.~(\ref{eq:Zh}) is not diffeomorphism invariant as the CFT is leaky into the bath. This is because the path integral measure and $S_{\text{matter}}[\psi;g^{0}]$ are invariant under Equ.~(\ref{eq:nontrivialTr}) but the interaction term transforms nontrivially
\begin{equation}
\begin{split}
    S_{\text{int}}[\psi,h;g^{0}]&\rightarrow S_{\text{int}}[\psi,h;g^{0}]+\sqrt{16\pi G_{N}}\int d^{2}x\sqrt{-g^{0}}\nabla_{\mu}\epsilon_{\nu}(x)T^{\mu\nu}_{\text{CFT}}\,,\\&=S_{\text{int}}[\psi,h;g^{0}]+\sqrt{16\pi G_{N}}\int dx\sqrt{-g^{0}_{\partial}}\epsilon_{\nu}(x)T^{n\nu}_{\text{CFT}}\,,\label{eq:deltaS}
    \end{split}
\end{equation}
where the upper index $n$ denotes the norm direction to the asymptotic boundary $x=\epsilon$ and we used the local conservation of the matter stress-energy tensor $\nabla_{\mu}T^{\mu\nu}_{\text{CFT}}=0$. The transparent boundary condition of the CFT leaking into the bath ensures that the resulting boundary flux term in Equ.~(\ref{eq:deltaS}) is nonzero. As a result, the path integral Equ.~(\ref{eq:Zh}) is not invariant under large diffeomorphisms due to the nonzero boundary flux.\footnote{In fact, as it is pointed out in  \cite{Geng:2025rov}, this observation already suggests that the diffeomorphism symmetry is spontaneously broken. We will review the relevant argument in Sec.~\ref{sec:flux}.} This result motivates us to embed the path integral Equ.~(\ref{eq:Zh}) into a manifestly diffeomorphism invariant description with Equ.~(\ref{eq:Zh}) as a result of gauge fixing. This embedding is naturally realized by introducing an auxiliary vector field $V^{\mu}(x)$ which, together with Equ.~(\ref{eq:nontrivialTr}), transforms under the diffeomorphism as
\begin{equation}
    V^{\mu}(x)\rightarrow V^{\mu}(x)-\epsilon^{\mu}(x)\,.
\end{equation}
The fully diffeomorphism invariant path integral is
\begin{equation}
    Z_{\text{full}}=\int D[h]D[\psi]D[V]e^{iS_{\text{matter}}[\psi;g^{0}]+iS_{\text{int}}[\psi,h;g^{0}]+i\sqrt{16\pi G_{N}}\int d^{2}x\sqrt{-g^{0}}\nabla_{\mu}V_{\nu}(x)T^{\mu\nu}_{\text{CFT}}}\,,\label{eq:Zfull}
\end{equation}
from which the original path integral Equ.~(\ref{eq:Zh}) can be obtained if one fixes the unitary gauge $V^{\mu}(x)=0$. 

The mass can be extracted from the low energy effective action of $h_{\mu\nu}$ and $V^{\mu}$, if one integrates out the transparent CFT fields $\psi$. Thus, we have
\begin{equation}
    S^{2}_{\text{eff}}[h,V]=2i\pi G_{N}\int d^{2}x\sqrt{-g^0}\int d^{2}y\sqrt{-g^0}(h_{\mu\nu}(x)+\nabla_{(\mu}V_{\nu)}(x))\langle T^{\mu\nu}_{\text{CFT}}(x)T^{\rho\sigma}_{\text{CFT}}(y)\rangle(h_{\rho\sigma}(y)+\nabla_{(\rho}V_{\sigma)}(y))\,,\label{eq:effbare}
\end{equation}
where $\nabla_{(\mu}V_{\nu)}=\nabla_{\mu}V_{\nu}+\nabla_{\nu}V_{\mu}$. We are only interested in the effective action at the quadratic level, which is the lowest nontrivial order in $G_{N}$, and we note that the effective action only depends on the diffeomorphism invariant combination
\begin{equation}
    \tilde h_{\mu\nu}=h_{\mu\nu}+\nabla_{\mu}V_{\nu}+\nabla_{\nu}V_{\mu}\,.\label{eq:invariant}
\end{equation}
Hence, we can consider $S^{2}_{\text{eff}}[0,V]$ and at the end replace $\nabla_{(\mu}V_{\nu)}$ by the invariant combination Equ.~(\ref{eq:invariant}). Note that the expression Equ.~(\ref{eq:effbare}) is divergent due to the singularity of the stress-energy tensor two-point function as $x\rightarrow y$. Thus, we should instead consider the renormalized stress-energy tensor two-point function which is obtained by vacuum substraction
\begin{equation}
    \langle T^{\mu\nu}_{\text{CFT}}(x)T^{\rho\sigma}_{\text{CFT}}(y)\rangle_{R}=\langle T^{\mu\nu}_{\text{CFT}}(x)T^{\rho\sigma}_{\text{CFT}}(y)\rangle-\langle T^{\mu\nu}_{\text{CFT}}(x)T^{\rho\sigma}_{\text{CFT}}(y)\rangle_{0}\,,
\end{equation}
where the vacuum stress-energy tensor two-point function is computed on the AdS$_2$ without the bath. The renormalized effective action is
\begin{equation}
    S^{2}_{\text{eff},R}[h,V]=2i\pi G_{N}\int d^{2}x\sqrt{-g^0}\int d^{2}y\sqrt{-g^0}(h_{\mu\nu}(x)+\nabla_{(\mu}V_{\nu)}(x))\langle T^{\mu\nu}_{\text{CFT}}(x)T^{\rho\sigma}_{\text{CFT}}(y)\rangle_{R}(h_{\rho\sigma}(y)+\nabla_{(\rho}V_{\sigma)}(y))\,.\label{eq:eff}
\end{equation}
The task is now reduced to the computation of the large distance behavior of the renormalized stress-energy tensor two-point function
\begin{equation}
    \langle T^{\mu\nu}_{\text{CFT}}(x)T^{\rho\sigma}_{\text{CFT}}(y)\rangle_{R}\,.\label{eq:TT}
\end{equation}
This two-point function can be computed if one notices that the AdS$_{2}$ metric Equ.~(\ref{eq:poincare}) is Weyl-equivalent to the flat space metric
\begin{equation}
    ds_{\text{flat}}^{2}=-dx^{+}dx^{-}\,,
\end{equation}
by the Weyl factor $\Omega=-\frac{4}{(x^{+}-x^{-})^{2}}$. The effect of this Weyl factor is to renormalize the parameter $\phi_{0}$ in the action Equ.~(\ref{eq:action}) as 
\begin{equation}
    \phi^{0}\rightarrow\phi^{0}+\frac{c G_{N}}{3}\,,
\end{equation}
which can be easily obtained using the CFT Weyl anomaly.\footnote{Note that in the usual discussion of the CFT Weyl anomaly one only has the bulk term $\frac{c}{48\pi}\int_{\mathcal{M}}d^{2}x\sqrt{-g}R$. In fact, the associated boundary term in our case can be easily obtained by making a Weyl transform for 2d Minkowski space which maps the left half to AdS$_{2}$ with the right half invariant. This Weyl transform has a $\theta$-function factor which contributes to the boundary term $\frac{c}{24\pi}\int_{\partial \mathcal{M}}K$ after taking the derivative.} Hence, we will ignore the Weyl transform of the stress-energy tensor hereafter and the stress-energy tensor two-point function is the same as in flat space with the following nonzero components\footnote{We note that the relative normalization factor between our convention of the stress-energy tensor and the standard CFT literature is $-\frac{1}{2\pi}$ (see \cite{DiFrancesco:1997nk} Equ.(5.40)).}
\begin{equation}
    \langle T_{++}(x^{+},x^{-})T_{++}(y^{+},y^{-})\rangle=\frac{c}{2(2\pi)^2(x^{+}-y^{+})^4}\,,\quad   \langle T_{--}(x^{+},x^{-})T_{--}(y^{+},y^{-})\rangle=\frac{c}{2(2\pi)^2(x^{-}-y^{-})^4}\,,\label{eq:TT}
\end{equation}
where we emphasize that we are imposing fully transparent boundary condition for the CFT on AdS$_{2}$. Therefore, the nonvanishing renormalized stress-energy tensor two-point function is
\begin{equation}
    \langle T_{++}(x^{+},x^{-})T_{--}(y^{+},y^{-})\rangle_{R}=-\frac{c}{2(2\pi)^2(x^{+}-y^{-})^4}\,,\quad   \langle T_{--}(x^{+},x^{-})T_{++}(y^{+},y^{-})\rangle_{R}=-\frac{c}{2(2\pi)^2(x^{-}-y^{+})^4}\,.\label{eq:TTR}
\end{equation}
The goal is to compare the large-distance behavior of the Equ.~(\ref{eq:TT}) with that of the Green's function of an AdS$_{2}$ vector field $A_{\mu}$ with mass squared as $2$. This vector field has the equation of motion
\begin{equation}
    \nabla_{\mu}(\nabla^{\mu}A_{\nu}-\nabla_{\nu}A^{\mu})-2A_{\nu}=0\,.
\end{equation}
Its Green's function has been calculated in \cite{Allen:1985wd} and it is given by
\begin{equation}
    G_{\mu \nu'}(X,X')=g_{\mu\nu'}(X,X')\big[(Z+1)(Z-1)\frac{d}{dZ}+Z\big]\gamma(Z)+n_{\mu}(X,X')n_{\nu'}(X,X')(Z-1)\big[(Z+1)\frac{d}{dZ}+1\big]\gamma(Z)\,, 
\end{equation}
where $X$ denotes $(x,t)$, $X'$ denotes $(y,t')$, $Z=\frac{x^2+x'^2-(t-t')^2}{2xx'}$ is the AdS$_{2}$ invariant distance and the function $\gamma(Z)$ is given by
\begin{equation}
\begin{split}
\gamma(Z)&=\frac{(-1)\Gamma[3]\Gamma[2]}{2^3 \pi \Gamma[4]}\frac{1}{2} \text{}_{2}F_{1}(3,2,4;\frac{2}{Z+1})\big(\frac{2}{Z+1}\big)^{3}\,\\&=-\frac{1}{8\pi}\big(\frac{2Z}{(Z+1)(Z-1)}+\log\frac{Z-1}{Z+1}\big)\,.
    \end{split}
\end{equation}
The covariant tensors are given by
\begin{equation}
    \begin{split}
n_{\mu}(X,X')=\frac{\partial_{\mu}Z}{\sqrt{Z^2-1}}\,,\quad n_{\nu'}(X,X')=\frac{\partial_{\nu'}Z}{\sqrt{Z^2-1}}\,,\quad g_{\mu\nu'}(Z,Z')=-\partial_{\mu}\partial_{\nu'}Z+\frac{\partial_{\mu}Z\partial_{\nu'}Z}{Z+
1}\,.\label{eq:Itensor}
    \end{split}
\end{equation}
Hence, in the large distance limit $Z\rightarrow\infty$ we have
\begin{equation}
    G_{\mu\nu'}(X,X')=\frac{1}{3\pi Z^2}\big(g_{\mu\nu'}+n_{\mu}n_{\nu'}\big)\,,
\end{equation}
and
\begin{equation}
    \nabla_{(\mu}G_{\nu)(\sigma'}(X,X')\nabla_{\rho')}=-\frac{3}{\pi Z^2}(g_{\mu\rho'}n_{\nu}n_{\sigma'}+g_{\mu\sigma'}n_{\nu}n_{\rho'}+g_{\nu\rho'}n_{\mu}n_{\sigma'}+g_{\nu\sigma'}n_{\mu}n_{\rho'}+4n_{\mu}n_{\nu}n_{\rho'}n_{\sigma'})\,.
\end{equation}
One can similarly decompose the stress-energy tensor two-point function in terms of the invariant tensors in Equ.~(\ref{eq:Itensor}) and the metric $g_{\mu\nu}$ and $g_{\rho'\sigma'}$ as
\begin{equation}
    \begin{split}
\langle T_{\mu\nu,\text{CFT}}(x)T_{\rho'\sigma',\text{CFT}}(y)\rangle=&\frac{c}{4(2\pi)^2(Z-1)^2}(g_{\mu\rho'}g_{\nu\sigma'}+g_{\mu\sigma'}g_{\nu\rho'}-g_{\mu\nu}g_{\rho'\sigma'})\,\\+&\frac{c}{2(2\pi)^2(Z-1)^2}(g_{\mu\rho'}n_{\nu}n_{\sigma'}+g_{\mu\sigma'}n_{\nu}n_{\rho'}+g_{\nu\rho'}n_{\mu}n_{\sigma'}+g_{\nu\sigma'}n_{\mu}n_{\rho'}+4n_{\mu}n_{\nu}n_{\rho'}n_{\sigma'})\,,                      \end{split}
\end{equation}
and the vacuum, i.e. reflective, stress-energy tensor two-point function as
\begin{equation}
    \begin{split}
\langle T_{\mu\nu,\text{CFT}}(x)T_{\rho'\sigma',\text{CFT}}(y)\rangle_{0}=&\frac{c}{4(2\pi)^2(Z+1)^2}(g_{\mu\rho'}g_{\nu\sigma'}+g_{\mu\sigma'}g_{\nu\rho'}-g_{\mu\nu}g_{\rho'\sigma'})\,.           \end{split}
\end{equation}
As a result, we have that in the large $Z$ limit
\begin{equation}
    \langle T_{\mu\nu,\text{CFT}}(x)T_{\rho'\sigma',\text{CFT}}(y)\rangle_{R}=-\frac{c\pi}{6(2\pi)^2}\nabla_{(\mu}G_{\nu)(\sigma'}(X,X')\nabla_{\rho')}\,.
\end{equation}

Now we are ready to compute
\begin{equation}
    \begin{split}
         S^{(2)}_{\text{eff},R}[0,V]&=2i\pi G_{N}\int d^{2}x\sqrt{-g^0}\int d^{2}y\sqrt{-g^0}\nabla_{(\mu}V_{\nu)}(x)\langle T^{\mu\nu}_{\text{CFT}}(x)T^{\rho\sigma}_{\text{CFT}}(y)\rangle_{R}\nabla_{(\rho}V_{\sigma)}(y)\,,\\&=-\frac{cG_{N}\pi^2}{3(2\pi)^2}i\int d^{2}x\sqrt{-g^0}\int d^{2}y\sqrt{-g^0}\nabla_{(\mu}V_{\nu)}(x)\nabla^{(\mu}G^{\nu)(\sigma'}(X,X')\nabla^{\rho')}\nabla_{(\rho'}V_{\sigma')}(y)\,,\\&=\frac{2cG_{N}\pi^2}{3(2\pi)^2}i\int d^{2}x\sqrt{-g^0}\int d^{2}y\sqrt{-g^0}V_{\nu}(x)\nabla_{\mu}\nabla^{(\mu}G^{\nu)(\sigma'}(X,X')\nabla^{\rho')}\nabla_{(\rho'}V_{\sigma')}(y)\,,\\&=\frac{4cG_{N}\pi^2}{3(2\pi)^2}\int d^{2}x\sqrt{-g^0}V_{\nu}(x)\big(\nabla^{\rho}\nabla_{\rho}V^{\nu}(x)+\nabla^{\rho}\nabla^{\nu}V_{\rho}(x)\big)\,,\\&=-\frac{2cG_{N}\pi^2}{3(2\pi)^2}\int d^{2}x\sqrt{-g^0}\big(\nabla^{\rho}V_{\nu}(x)+\nabla_{\nu}V^{\rho}(x)\big)\big(\nabla_{\rho}V^{\nu}(x)+\nabla^{\nu}V_{\rho}(x)\big)\,,
         \label{eq:effV}
    \end{split}
\end{equation}
where we used the following formula for the Green's function
\begin{equation}
    \nabla_{\mu}\nabla^{\mu}G^{\nu\sigma'}(X,X')-\nabla_{\mu}\nabla^{\nu}G^{\mu\sigma'}(X,X')-2 G^{\nu\sigma'}(X,X')=-i\frac{g^{\nu\sigma'}}{\sqrt{-g}}\delta^{2}(X-X')\,.\label{eq:Geom}
\end{equation}
We note that in fact a few terms are missed in Equ.~(\ref{eq:effV}) when we are integrating by parts as from Equ.~(\ref{eq:Geom}) we can see that the Green's function $G^{\mu\nu}$ is divergenceless. Therefore, we in fact have
\begin{equation}
    \begin{split}
         S^{(2)}_{\text{eff},R}[0,V]&=-\frac{2cG_{N}\pi^2}{3(2\pi)^2}\int d^{2}x\sqrt{-g^0}\big(\nabla^{\rho}V_{\nu}(x)+\nabla_{\nu}V^{\rho}(x)\big)\big(\nabla_{\rho}V^{\nu}(x)+\nabla^{\nu}V_{\rho}(x)\big)\,,\\&=-\frac{cG_{N}}{6}\int d^{2}x\sqrt{-g^0}\big(\nabla^{\rho}V_{\nu}+\nabla_{\nu}V^{\rho}\big)\big(\nabla_{\rho}V^{\nu}+\nabla^{\rho}\nabla^{\nu}V_{\rho}\big)-4\nabla_{\nu}V^{\nu}\nabla_{\rho}V^{\rho}\,.
         \label{eq:effVfinal}
    \end{split}
\end{equation}
As a result, we have
\begin{equation}
    S^{(2)}_{\text{eff},R}[h,V]=-\frac{M^2}{4}\int d^{2}x\sqrt{-g^0}\big(\tilde{h}_{\mu\nu}\tilde{h}^{\mu\nu}-\tilde{h}\tilde{h}\big)\,,\label{eq:mass}
\end{equation}
which is the St\"{u}ckelberg mass term in the Fierz-Pauli form and the mass squared is
\begin{equation}
    M^2=\frac{2}{3}\frac{c G_{N}}{l_{AdS}^{2}}\,,\label{eq:massresult}
\end{equation}
where we restored the dependence on the AdS length scale.

\subsection{The Mass versus Energy Dissipation}\label{sec:massvsdiss}
It has been suggested that the analogy of the higher dimensional graviton mass in 2d systems, like the one we considered in Sec.~\ref{sec:direct}, is the characteristic scale of energy dissipation from AdS to the bath.\footnote{This is pointed out by Henry W. Lin \cite{Almheiri:2021up} in 2022 and Emanuel Katz in 2023. We thank Henry W. Lin for bringing up this point and we thank Emanuel Katz and Liam A. Fitzpatrick for the relevant discussions. } This is a reasonable guess. In this section, we will demonstrate that generically these two scales are rather different from each other. Furthermore, we will see that  from the gravitational point of view these two scales correspond to rather different physics.

\subsubsection{A Toy Quantum Mechanics Model}\label{sec:toy}
Let's first consider a toy model of energy dissipation into a bath where the bath-induced mass and the characteristic scale of energy dissipation can be at best accidentally close. We will comment on the relationship between this toy model and the graviton mass in the island model.

We consider the following system of coupled harmonic oscillators
\begin{equation}
    L=\frac{1}{2}M_{0}\dot{q}^2(t)-\frac{1}{2}\omega^2_{0}M_{0}q^2(t)+\sum_{i}\big(\frac{1}{2}m_{i}\dot{x}_{i}^2(t)-\frac{1}{2}\omega_{i}^2m_{i}x_{i}^{2}(t)\big)-q(t)\sum_{i}g_{i}x_{i}(t)\,,\label{eq:toyaction}
\end{equation}
which is called the Caldeira–Leggett model \cite{Breuer:2007juk,Weiss:2021uhm} and where the harmonic oscillators $x_{i}(t)$ are modeling the bath for the ``system harmonic oscillator" $q(t)$. When the system and the bath are decoupled, i.e. $g_{i}=0$, the mass of the system is $M_{0}$. However, when the system and the bath are coupled, one can see that the dynamics of the system is modified by integrating out the bath. In the equilibrium states, the bath-induced Lagrangian of the system is given by
\begin{equation}
\begin{split}
    \delta L=&i\frac{1}{2}\int dt'q(t)q(t')\sum_{i}g_{i}^2 \langle T x_{i}(t)x_{i}(t')\rangle\,,\\=&-\frac{i}{2}\sum_{i}g_{i}^{2}\int dt'\int \frac{d\omega}{2\pi} q(t)q(t')e^{-i\omega(t-t')}\frac{1}{i(\omega^2-\omega_{i}^2+i\epsilon)m_{i}}\,,\\=&-\frac{i}{2\pi}\sum_{i}g_{i}^{2}\int_{-\infty}^{t} dt' q(t)q(t')e^{-i(\omega_{i}-i\epsilon)(t-t')}\frac{-2\pi i}{4i\omega_{i}m_{i}}+\frac{i}{2\pi}\sum_{i}g_{i}^{2}\int_{t}^{\infty} dt' q(t)q(t')e^{i(\omega_{i}-i\epsilon)(t-t')}\frac{2\pi i}{4i\omega_{i}m_{i}}\,,\\=&\frac{i}{2\pi}\sum_{i}g_{i}^{2}\int_{0}^{\infty} dt' q(t)q(t-t')e^{-i(\omega_{i}-i\epsilon)t'}\frac{2\pi i}{4i\omega_{i}m_{i}}+\frac{i}{2\pi}\sum_{i}g_{i}^{2}\int_{-\infty}^{0} dt' q(t)q(t-t')e^{i(\omega_{i}-i\epsilon)t'}\frac{2\pi i}{4i\omega_{i}m_{i}}\,,
    \end{split}
\end{equation}
which can be then systematically expanded as a derivative expansion. We are interested in the low derivative terms. The relevant terms are
\begin{equation}
\begin{split}
     \delta L_{0}=&\frac{i}{2\pi}\sum_{i}g_{i}^{2}\int_{0}^{\infty} dt' q(t)q(t)e^{-i(\omega_{i}-i\epsilon)t'}\frac{2\pi i}{4i\omega_{i}m_{i}}+\frac{i}{2\pi}\sum_{i}g_{i}^{2}\int_{-\infty}^{0} dt' q(t)q(t)e^{i(\omega_{i}-i\epsilon)t'}\frac{2\pi i}{4i\omega_{i}m_{i}}\,,\\=&\sum_{i}\frac{g_{i}^2 i}{2\omega_{i}m_{i}}\int_{0}^{\infty}dt'e^{-i(\omega-i\epsilon)t'}q^{2}(t)\,\,,\\=&q^{2}(t)\sum_{i}\frac{g_{i}^2}{2\omega_{i}^2m_{i}}\,.
     \end{split}
\end{equation}
\begin{equation}
\begin{split}
     \delta L_{1}=&\frac{i}{2\pi}\sum_{i}g_{i}^{2}\int_{0}^{\infty} dt' q(t)\dot{q}(t)e^{-i(\omega_{i}-i\epsilon)t'}\frac{-2\pi t' i}{4i\omega_{i}m_{i}}+\frac{i}{2\pi}\sum_{i}g_{i}^{2}\int_{-\infty}^{0} dt' q(t)\dot{q}(t)e^{i(\omega_{i}-i\epsilon)t'}\frac{-2\pi t' i}{4i\omega_{i}m_{i}}\,,\\=0\,.
     \end{split}
\end{equation}
\begin{equation}
\begin{split}
     \delta L_{2}=&\frac{i}{2\pi}\sum_{i}g_{i}^{2}\int_{0}^{\infty} dt' q(t)\dot{q}(t)e^{-i(\omega_{i}-i\epsilon)t'}\frac{\pi t'^2 i}{4i\omega_{i}m_{i}}+\frac{i}{2\pi}\sum_{i}g_{i}^{2}\int_{-\infty}^{0} dt' q(t)\dot{q}(t)e^{i(\omega_{i}-i\epsilon)t'}\frac{\pi t'^2 i}{4i\omega_{i}m_{i}}\,,\\=&-\sum_{i}\frac{g_{i}^2}{2\omega_{i}^4 m_{i}}q(t)\Ddot{q}(t)\,.\label{eq:masscorrection}
     \end{split}
\end{equation}
As a result, we have the quantum corrected mass and frequency for the system
\begin{equation}
    M=M_{0}+\sum_{i}\frac{g_{i}^2}{\omega_{i}^4 m_{i}}\,,\quad M\omega_{\text{new}}^2=M_{0}\omega_{0}^2-\sum_{i}\frac{g_{i}^2}{\omega_{i}^2 m_{i}}\,,\label{eq:renormalization}
\end{equation}
and the dissipative nature of the system is not manifest as we are only considering equilibrium states.

The dissipative nature of the dynamics of the system induced by the bath will manifest if one considers nonequilibrium processes. Let's consider the process of turning on the coupling between the bath and the system at $t=0$, i.e. we take the coupling constants $g_{i}$ in Equ.~(\ref{eq:toyaction}) to be
\begin{equation}
    g_{i}(t)=\Theta(t)g_{i}\,.
\end{equation}
The equations of motion of the bath degrees of freedom for $t>0$ are
\begin{equation}
    m_{i}\Ddot{x}_{i}(t)+m_{i}\omega_{i}^2 x_{i}(t)+g_{i}q(t)=0\,.
\end{equation}
Thus, one can write the bath degrees of freedom using the retarded Green's function as
\begin{equation}
    x_{i}(t)=-g_{i}\int_{0}^{t} dt' G_{i}^{R}(t,t') q(t')=-\frac{g_{i}}{m_{i}\omega_{i}}\int_{0}^{t}dt'\sin\omega_{i}(t-t')q(t')\,,
\end{equation}
and plugging this expression to the equation of motion of the system we have
\begin{equation}
    M_{0}\Ddot{q}(t)+M_{0}\omega_{0}^2 q(t)-\sum_{i}\frac{g_{i}^2}{m_{i}\omega_{i}}\int_{0}^{t}dt'\sin\omega_{i}(t-t')q(t')=0\,,
\end{equation}
which is equivalent to
\begin{equation}
    M_{0}\Ddot{q}(t)+(M_{0}\omega_{0}^2-\sum_{i}\frac{g_{i}^2}{m_{i}\omega_{i}^2}) q(t)+\sum_{i}\frac{g_{i}^2}{m_{i}\omega_{i}^2}\int_{0}^{t}dt'\cos\omega_{i}(t-t')\dot{q}(t')=0\,,
\end{equation}
where we assumed that $q(0)=0$. The dissipative dynamics manifests if one considers an Ohmic bath for which $\frac{g_{i}^2}{m_{i}\omega_{i}^2}=M_{0}\gamma\omega_{B}$ and $\omega_{n}=2\pi n \omega_{B}$ with $n\geq 1$. This implies that for $|t-t'|\ll\frac{1}{\omega_{B}}$ we have
\begin{equation}
    \sum_{i} \frac{g_{i}^2}{m_{i}\omega_{i}^2}\cos\omega_{i}(t-t')=M_{0}\gamma\delta(t-t')\,,
\end{equation}
and the equation of motion becomes
\begin{equation}
    M_{0}\Ddot{q}(t)+(M_{0}\omega_{0}^2-\sum_{i}\frac{g_{i}^2}{m_{i}\omega_{i}^2}) q(t)+M_{0}\gamma\dot{q}(t)=0\,.
\end{equation}
Thus, the characteristic energy scale of energy dissipation is
\begin{equation}
    \Lambda_{\text{dissipation}}=\gamma\,,
\end{equation}
and the shift of the mass when the couplings $g_{i}$ are weak, due to Equ.~(\ref{eq:renormalization}), is
\begin{equation}
    \delta M=\gamma \frac{M_{0}}{24\omega_{B}}\,. 
\end{equation}
As a result, we can see that these two scales, $\Lambda_{\text{dissipation}}$ and $\delta M$, are not the same in general but they can be accidentally the same if one tunes the ratio $\frac{M_{0}}{\omega_{B}}$. Though, they scale in the same manner with the coupling strength in the weak dissipative coupling limit which is accidentally the case by dimensional analysis.

However, this toy example so far is quite different from the island model even in spirit. This is because in the island model the graviton mass is induced from the coupling between the graviton and leaky matter fields. In the above example in Equ.~(\ref{eq:toyaction}), the system degree of freedom $q(t)$ is in fact modeling the leaky matter fields. Thus, to enrich the toy model in Equ.~(\ref{eq:toyaction}) such that it could capture the physics in the island model, we have to introduce another harmonic oscillator $h(t)$ in the system that is coupled to the leaky mode $q(t)$ by a coupling term
\begin{equation}
    \delta L_{\text{int}}=g_h h(t)q^2(t)\,,
\end{equation}
with the coupling constant $g^2_{h}$ in analogy with the Newton's constant $G_{N}$. The correction to the mass of $h(t)$ can be similarly exacted as in Equ.~(\ref{eq:masscorrection}) but with a squared time-ordered propagator. The result can be read off as
\begin{equation}
    M_{h}=M_{h,0}+\frac{g_{h}^2}{4\omega_{\text{new}}^5 M^2}\,.
\end{equation}
So the leakiness induced mass correction is
\begin{equation}
    \delta M_{h}=\gamma\frac{g_{h}^2M_{0}^3}{32(M\omega_{\text{new}})^5 \omega_{B}}\,,
\end{equation}
which is rather different from the dissipative energy scale $\gamma$. Hence, this toy model calculation already suggests that the graviton mass in the island model and the characteristic energy scale associated with the energy dissipation into the bath are rather different scales. One of them describes equilibrium physics and the other characterizes the dynamics in nonequilibrium states. 

Thus, one naturally expects that the 2d analog of the graviton mass in the island model and the nonzero energy dissipative scale are rather different scales. In the next subsection, we will confirm this expectation. We will extract the energy dissipation scale in the leaky CFT coupled JT model and we will see that the answer is parametrically different from the mass scale we calculated in Sec.~\ref{sec:direct}. We will also see that from the gravitational point of view these two scales are associated with rather different physics.

\subsubsection{The CFT Coupled JT Model}\label{sec:CFTdissip}

In this subsection, we will extract the characteristic energy scale associated with the energy dissipation into the bath in the CFT coupled JT model. As we have discussed in Sec.~\ref{sec:toy}, in order to extract this scale we have to consider nonequilibrium processes.

Before we trigger and study the nonequilibrium process, let's first outline some general results in JT gravity. The equations of motion of the matter coupled JT gravity, whose action is in the same form as Equ.~(\ref{eq:action}), are
\begin{equation}
    R=-2\,,\quad 
    8\pi G_{N} T_{\mu\nu,\text{m}}=-\nabla_{\mu}\nabla_{\nu}\phi+g_{\mu\nu}\nabla^{2}\phi-g_{\mu\nu}\phi\,.\label{eq:Einstein}
\end{equation}
The first equation localizes the spacetime geometry at the quantum level to be empty AdS$_{2}$ which we use the Poincar\'{e} patch as in Equ.~(\ref{eq:poincare}).\footnote{We don't consider higher topologies in the Lorenzian signature.} However, the dynamics of JT gravity is nontrivial due to the fact that we don't impose the normalizable boundary condition for the dilaton and the metric when we send the boundary cutoff $\epsilon$ to zero as in the usual study of AdS/CFT correspondence \cite{Maldacena:1997re,Gubser:1998bc,Witten:1998qj}. Instead, one chooses the non-normalizable boundary condition for the dilaton and fixed the boundary metric using the boundary time-coordinate $u$ as
\begin{equation}
    ds^{2}|_{\partial}=-\frac{t'^2(u)-x'^2(u)}{x^2(u)}=-\frac{du^2}{\epsilon^2}\,,\quad \phi_{\partial}=\frac{\phi_{r}}{\epsilon}\,,\label{eq:bdy}
\end{equation}
with $\epsilon\rightarrow0$ and $\phi_{r}$ a fixed number \cite{Almheiri:2014cka,Engelsoy:2016xyb,Jensen:2016pah,Maldacena:2016upp,Nayak:2018qej,Yang:2018gdb}.\footnote{In fact, it is rather hard to make sense of this this boundary condition at a fine-grained level \cite{Geng:2022tfc}. But it is okay at the level of an EFT so we will not be bothered by this issue hereafter.} Let's denote $t=f(u)$, so we have from the first equation of Equ.~(\ref{eq:bdy}) the boundary coordinates
\begin{equation}
    t_{\partial}=f(u)\,,\quad x_{\partial}=\epsilon f'(u)+\mathcal{O}(\epsilon^3)\,.
\end{equation}
Thus, we can extend the boundary coordinates into the bulk as
\begin{equation}
    x^{+}=t+x=f(u+y)\,,\quad x^{-}=t-x=f(u-y)\,,
\end{equation}
with the boundary at $y=\epsilon$. We will call $(u,y)$ as the \textit{physical coordinates} and $(t,x)$ as the \textit{fictitious coordinates}. Thus, the spacetime metric is fixed in the fictitious coordinates but not under the physical coordinates
\begin{equation}
    ds^2=-\frac{4dx^{+}dx^{-}}{(x^{+}-x^{-})^2}=-\frac{4f'(y^{+})f'(y^{-})}{\big(f(y^{+})-f(y^{-})\big)^2}dy^{+}dy^{-}\,,\quad\text{with } y^{\pm}=u\pm y\,.
\end{equation}
Solving the second equation in Equ.~(\ref{eq:Einstein}), we have \cite{Engelsoy:2016xyb}
\begin{equation}
    \phi(x^{+},x^{-})=\frac{a}{x^{+}-x^{-}}\Big(1-b(x^{+}+x^{-})-dx^{+}x^{-}-I_{+}(x+,x-)-I_{-}(x^{+},x^{-})\Big)\,.\label{eq:dilaton}
\end{equation}
where we defined
\begin{equation}
    \begin{split}
I_{+}(x^{+},x^{-})&=\frac{8\pi G_{N}}{a}\int_{x^{+}}^{\infty} ds(s-x^{+})(s-x^{-})T_{++,\text{m}}(s)\,,\\ I_{-}(x^{+},x^{-})&=\frac{8\pi G_{N}}{a}\int_{-\infty}^{x^{-}}ds(s-x^{+})(s-x^{-})T_{--,\text{m}}(s)\,,
    \end{split}
\end{equation}
with the constants $a$, $b$ and $d$ chosen such that the second condition of Equ.~(\ref{eq:bdy}) is obeyed. In fact, the constants $b$ and $d$ should be zero and $a=2\phi_{r}$. This is because when the matter source is zero the physical coordinates and the fictitious coordinates are the same and the second equation in Equ.~(\ref{eq:bdy}) would be satisfied. Thus, from Equ.~(\ref{eq:dilaton}) we have the following result near the boundary of the AdS$_{2}$:
\begin{equation}
    \phi_{\partial}=\frac{\phi_{r}}{\epsilon f'(u)}\big(1-I_{+}(f(u))-I_{-}(f(u))\big)+\mathcal{O}(1)\,.
\end{equation}
Hence, the second condition of Equ.~(\ref{eq:bdy}) implies the following constraint
\begin{equation}
\begin{split}
    f'(u)&=1-I_{+}(f(u))-I_{-}(f(u))\,,\\&=1-\frac{4\pi G_{N}}{\phi_{r}}\int_{f(u)}^{\infty} ds (s-f(u))^{2}T_{++,\text{m}}(s)-\frac{4\pi G_{N}}{\phi_{r}}\int_{-\infty}^{f(u)} ds (s-f(u))^{2}T_{--,\text{m}}(s)\,,
    \end{split}
\end{equation}
taking three derivatives with respect to $f(u)$ one has
\begin{equation}
    \frac{d^{3}}{df(u)^3}f'(u)=\frac{8\pi G_{N}}{\phi_{r}}\big(T_{++,\text{m}}(f(u))-T_{--,\text{m}}(f(u))\big)\,,
\end{equation}
which is equivalent to
\begin{equation}
    \frac{d}{du}\{f(u),u\}=\frac{8\pi G_{N}}{\phi_{r}}f'^2(u) \big(T_{++,\text{m}}(f(u))-T_{--,\text{m}}(f(u))\big)\,,
\end{equation}
where $\{f(u),u\}$ denotes the Schwarzian of the function $f(u)$.

We will use the coupling protocol in \cite{Almheiri:2019psf} to trigger the nonequilibrium process by suddenly turning on the coupling between the AdS$_{2}$ and a bath. We take the matter fields in AdS$_{2}$ to be conformal field theories and in the bath we have the same conformal field theory. Before the coupling is turned on, they are both in the vacuum state and the coupling enables the energy-momentum to be completely transparent between the AdS$_{2}$ and the bath. The calculation in \cite{Almheiri:2019psf} suggests that after the coupling is turned on, we have the boundary stress-energy flux
\begin{equation}
    T_{++}(u)=0\,,\quad T_{--}(u)=\frac{c}{24\pi}\frac{1}{f'^2 (u)}\{f(u),u\}\,.
\end{equation}
Thus, after the coupling is turned on we have 
\begin{equation}
    \frac{d}{du}\{f(u),u\}=-\frac{cG_{N}}{3\phi_{r}}\{f(u),u\}\,.
\end{equation}
As a result, we have the characteristic energy scale associated with the energy dissipation into the bath as
\begin{equation}
    \Lambda_{\text{dissipation}}=\frac{cG_{N}}{3\phi_{r}}\,.\label{eq:JTdiss}
\end{equation}
We note that this energy scale is parametrically different from the mass scale we extracted in the equilibrium states in Equ.~(\ref{eq:massresult}), and it is of a higher order in the Newton's constant $G_{N}$, unless one fine tunes the parameter $\phi_{r}$.

Having obtained both the mass Equ.~(\ref{eq:massresult}) and the dissipation scale Equ.~(\ref{eq:JTdiss}) and noticed that they are parametrically different from each other, we would like to further articulate the difference between their physical meanings.

In Sec.~\ref{sec:direct} we started with the action of JT gravity coupled with leaky conformal matter fields in Equ.~(\ref{eq:action}) and integrated out the conformal matter fields in the gravitational path integral. The result of this analysis is that the backreaction of the matter fields to the gravity sector to the leading nontrivial order in $G_{N}$ is an induced mass term Equ.~(\ref{eq:mass}). To see the implications of this additional term, let's turn on some other matter fields which though obey reflective boundary conditions. In the absence of the leaky CFT, using the second equation in Equ.~(\ref{eq:Einstein}), the energy of the matter fields can be expressed as
\begin{equation}
    H_{\text{matter}}=-\int dx \sqrt{-g} T^{0}_{0,\text{matter}}=\int dx\sqrt{-g}\Big(-\frac{1}{8\pi G_{N}}g^{xx}\nabla_{x}\nabla_{x}\phi+\frac{1}{8\pi G_{N}}\phi\Big)\,,
\end{equation}
which can be further simplified as
\begin{equation}
    H_{\text{matter}}=-\frac{1}{8\pi G_{N}}\int dx \partial_{x}\Big(\partial_{x}\phi+\frac{\phi}{x}\Big)=\frac{1}{8\pi G_{N}}\Big(\partial_{x}\phi+\frac{\phi}{x}\Big)\big\vert_{x=x_{\partial}}\equiv H_{\text{ADM}}\,.\label{eq:rigidGauss}
\end{equation}
The result Equ.~(\ref{eq:rigidGauss}) is nothing but the gravitational Gauss' law which states that the energy of a gravitational spacetime can be measured on its boundary using perturbative gravitational effects \cite{Geng:2023zhq}. The gravitational Gauss' law holds due to the long-range nature of the standard massless gravity theories. On the other hand, in the presence of the leaky CFT, the equations of motion of the induced theory after integrating out the leaky conformal matter fields are
\begin{equation}
    R=-2\,,8\pi G_{N}T_{\mu\nu,\text{matter}}=-\nabla_{\mu}\nabla_{\nu}\phi+g_{\mu\nu}\nabla^{2}\phi-g_{\mu\nu}\phi+\frac{8\pi G_{N}}{\sqrt{16\pi G_{N}}} M^2 (\tilde{h}_{\mu\nu}-g_{\mu\nu}\tilde{h}_{\mu\nu})\,,\label{eq:Einsteininduced}
\end{equation}
which is in fact a linearized equation of motion to the first nontrivial order in $G_{N}$. Thus, we have
\begin{equation}
\begin{split}    
H_{\text{matter}}&=H_{\text{ADM}}-\frac{M^2}{\sqrt{16\pi G_{N}}}\int dx\sqrt{-g}(\tilde{h}^{0}_{0}-\tilde{h}^{\mu}_{\mu})=H_{\text{ADM}}+\frac{M^2}{\sqrt{16\pi G_{N}}}\int dx \tilde{h}_{xx}\\&=H_{\text{ADM}}+\frac{M^2}{16\pi G_{N}}\int dx (\delta g_{xx}+2\sqrt{16\pi G_{N}}\nabla_{x}V_{x})\,,\label{eq:massivegauss}
\end{split}
\end{equation}
where $\delta g_{\mu\nu}$ is the fluctuation of the metric due to the matter backreaction. Thus, we can see that it is no longer true that any energetic excitations inside a spacetime can be detected on its boundary using perturbative gravitational effects. This is because energy can be hidden inside the bulk due to the fact that the gravitational theory is no longer long-ranged \cite{Geng:2021hlu,Geng:2025rov}. A more careful treatment of this point will be provided in Sec.~\ref{sec:necc}.

In summary, we can conclude that both the parametric values and the physics associated with mass Equ.~(\ref{eq:massresult}) and the energy dissipation scale Equ.~(\ref{eq:JTdiss}) are rather different. The former enables information to be localized inside a gravitational spacetime, such that it is not detectable on the asymptotic boundary of this gravitational universe. The latter provides a characteristic time scale for the system to approach equilibrium again after some external perturbation. Entanglement islands exist in both the equilibrium states and the nonequilibrium states as long as the AdS and the bath are coupled. Furthermore, a consistency condition for the island is that information should be able to be localized inside the island from being detected by the asymptotic boundary \cite{Geng:2021hlu}. This consistency condition is satisfied in the island model due to the mass Equ.~(\ref{eq:massresult}). Thus, we can see that the dissipative scale Equ.~(\ref{eq:JTdiss}) is in fact not relevant to islands but the mass Equ.~(\ref{eq:massresult}) is essential. From a more practical point of view, the dissipation scale Equ.~(\ref{eq:JTdiss}) is parametrically much lower than the mass Equ.~(\ref{eq:massresult}) in physically relevant situations, i.e. weakly coupled gravity with only a few matter fields. Therefore, the energy dissipation process is in fact extremely slow and we are safe to treat the system adiabatically at each instant. We can thus look for islands at each instant. So the energy dissipation scale is not relevant to island both physically and practically.

\subsection{A Holographic Model}\label{sec:holographic}
We have seen from Sec.~\ref{sec:direct} and Sec.~\ref{sec:CFTdissip} that in the island model there is a bath coupling induced mass Equ.~(\ref{eq:massresult}), the induced mass term for the gravitational action is in the St\"{u}ckelberg form and this correction to the gravitational action enables information to be localized inside the island from being able to be detected by the asymptotic boundary. In fact, there is an interesting model where all of these observations are manifest at the intuitive level. This model also has important implications to the emergence of spacetime from quantum entanglement \cite{Geng:2025rov}. In this subsection, we will revisit this model from a lower dimensional perspective, to stay close to the spirit of this section. Discussions of the implications of this model will be presented in Sec.~\ref{sec:necc}.

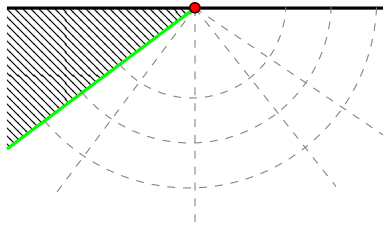
\begin{figure}
\begin{centering}
\begin{tikzpicture}[scale=1]
\draw[-,very thick,black!100] (-2.5,0) to (0,0);
\draw[-,very thick,black!100] (0,0) to (2.5,0);
\draw[pattern=north west lines,pattern color=gray!200,draw=none] (0,0) to (-2.5,-1.875) to (-2.5,0) to (0,0);
%violet!50
\draw[-,dashed,color=black!50] (0,0) to (-1.875,-2.5); 
\draw[-,dashed,color=black!50] (0,0) to (0,-2.875);
\draw[-,dashed,color=black!50] (0,0) to (1.875,-2.375); 
\draw[-,dashed,color=black!50] (0,0) to (2.5,-1.6875); 
\draw[-,very thick,color=green!!50] (0,0) to (-2.5,-1.875);
%\draw[-] (-0.75,0) arc (180:217.5:0.75);
%\node at (-0.5,0.2) {$\mu$};
%\draw[-] (1.5,0) arc (0:-5.25:1.5);
\node at (0,0) {\textcolor{red}{$\bullet$}};
\node at (0,0) {\textcolor{black}{$\circ$}};
\draw[-,dashed,color=black!50] (-2,-1.5) arc (-140:-2:2.5);
\draw[-,dashed,color=black!50] (-1.5,-1.125) arc (-140:-2:1.875);
\draw[-,dashed,color=black!50] (-1,-0.75) arc (-140:-2:1.25);
\end{tikzpicture}
\caption{\small A constant time slice of an AdS$_{d+1}$ with a Karch-Randall brane. The green surface denotes the brane and it has AdS$_{d}$ geometry. The gray-shaded region behind the brane is cutoff. The red dot is the codimension one submanifold of the bulk asymptotic boundary (the thick black line) where it intersects the brane. This submanifold is also called the \textit{defect} in the literature. }
\label{pic:KR}
\end{centering}
\end{figure}

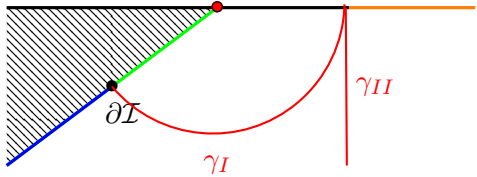
\begin{figure}
\begin{centering}
\begin{tikzpicture}[scale=1.4]
\draw[-,very thick,black!100] (-2,0) to (0,0);
\draw[-,very thick,black!100] (0,0) to (1.25,0);
\draw[-,very thick,orange] (1.25,0) to (2.45,0);
\draw[pattern=north west lines,pattern color=gray!200,draw=none] (0,0) to (-2,-1.5) to (-2,0) to (0,0);
%violet!50
\draw[-,very thick,color=green!!50] (0,0) to (-2,-1.5);
\node at (0,0) {\textcolor{red}{$\bullet$}};
\node at (0,0) {\textcolor{black}{$\circ$}};
\draw[-,very thick,blue] (-1,-0.75) to (-2,-1.5);
\node at (-1,-0.75) {\textcolor{black}{$\bullet$}};
\node at (-0.9,-1) {\textcolor{black}{$\partial\mathcal{I}$}};
\draw[-,thick,color=red] (-1,-0.75) arc (-140:-1:1.25);
\draw[-,thick,color=red] (1.22,0) to (1.23,-1.5);
\node at (0,-1.49) {\textcolor{red}{$\gamma_{I}$}};
\node at (1.5,-0.75) {\textcolor{red}{$\gamma_{II}$}};
\end{tikzpicture}
\caption{\small A demonstration of the construction of entanglement island in the Karch-Randall braneworld. The orange line denotes the bath subregion $R$ with blue region on the brane denoting its entanglement island $\mathcal{I}$. The red surface $\gamma_{I}$ connecting $\partial R$ and $\partial\mathcal{I}$ is a minimal area surface. $\gamma_{I}$ is determined by minimizing its area also with respect to its ending point $\partial \mathcal{I}$ on the brane which correspondence to the $\min_{ \mathcal{I}}$ in the island formula Equ.~(\ref{eq:islandformula}). $\gamma_{I}$ and $\gamma_{II}$ are called the Ryu-Takayanagi surfaces.}
\label{pic: KRisland}
\end{centering}
\end{figure}

This model is the so-called Karch-Randall braneworld \cite{Karch:2000ct,Karch:2000gx,Geng:2023qwm}. The Karch-Randall braneworld describes the physics of an AdS$_{d}$ end-of-the-world brane embedded in an ambient gravitational AdS$_{d+1}$ spacetime (see Fig.~\ref{pic:KR}). This model has a nice holographic interpretation as the dual of a quantum gravitational theory on AdS$_{d}$ coupled with a nongravitational bath with transparent matter fields propagating between them. From the ambient AdS$_{d+1}$ perspective, one can think of the AdS$_{d}$ as the Karch-Randall brane and the bath as the leftover asymptotic boundary of the AdS$_{d+1}$. This is a natural model to construct and study entanglement islands in both lower \cite{Almheiri:2019hni} and higher dimensions \cite{Almheiri:2019psy,Geng:2020qvw,Chen:2020uac,Chen:2020hmv}. This is because islands can be easily searched using holographic tools \cite{Geng:2024xpj} (see Fig.~\ref{pic: KRisland}). More interestingly, the essential physics in the island model are nicely geometrized in this model \cite{Geng:2025rov}. 

Let's revisit the relevant physics in this setup using a lower-dimensional case, with an AdS$_{2}$ Karch-Randall brane in an AdS$_{3}$ ambient spacetime. The action of this system is 
\begin{equation}
\begin{split}
    S=&\frac{1}{16\pi G_{N}}\int_{\text{bulk}} d^{3}x\sqrt{-g}\big(R-2\Lambda\big)+\frac{1}{8\pi G_{N}}\int_{\text{brane}} d^{2}x\sqrt{-h}(K+T)\,,\label{eq:KRaction}
    \end{split}
\end{equation}
where the bulk cosmological constant is given as 
\begin{equation}
    \Lambda=-\frac{1}{L_{AdS}^2}\,,
\end{equation}
with $L_{AdS}$ as the bulk AdS$_3$ length scale, $T_{B}$ is the brane tension and we ignored the Gibbons-Hawking term near the bulk asymptotic boundary for convenience. The essential calculations have been carried out in detail in Section 4 of \cite{Geng:2025rov} for generic dimensions. Here we outline the important steps to see the emergence of the St\"{u}ckelberg mass term Equ.~(\ref{eq:mass}) for the dual gravitational theory on AdS$_{2}$ and the physical meaning of the Goldstone vector field in this setup. 

For generality, let's take the bulk geometry to be of the form
\begin{equation}
\frac{ds^2}{L_{AdS}^2}=d\rho^{2}+e^{2A(\rho)}\bar{g}_{ij}(x)dx^{i}dx^{j}\,,\label{eq:metric}
\end{equation}
where for AdS$_{3}$ we have $e^{A(\rho)}=\cosh\rho$, $\bar{g}_{ij}(x)$ is the metric for a unit length AdS$_{2}$, the brane resides along a slice $\rho=\rho_{B}$ with $\rho_{B}$ determined by the brane tension $T$ as $T=\frac{1}{L_{AdS}}\tanh\rho_{B}$ \cite{Geng:2023qwm} and the coordinate $\rho$ lives in the regime $\rho\in[\rho_{B},\infty)$. The linearized Einstein's equation in the AdS$_{3}$ bulk is
\begin{equation}
    \frac{1}{2}g_{\mu\nu}\nabla_{\alpha}\nabla_{\beta}h^{\alpha\beta}-\frac{1}{2}g_{\mu\nu}\Box h+\frac{1}{2}\nabla_{\mu}\nabla_{\nu}h-\frac{1}{2}\nabla_{\alpha}\nabla_{\mu}h_{\nu}^{\alpha}-\frac{1}{2}\nabla_{\alpha}\nabla_{\nu}h_{\mu}^{\alpha}+\frac{1}{2}\Box h_{\mu\nu}+(g_{\mu\nu}h-2h_{\mu\nu})\,,\label{eq:linearizedEF}
\end{equation}
where $g_{\mu\nu}$ denotes the bulk background metric and we have set the bulk AdS$_3$ length scale to one. The fundamental field in the above equation of motion is the metric fluctuation, i.e. the graviton field, $h_{\mu\nu}$ and all upper indices are induced by contracting the lower ones with the inverse background metric $g^{\mu\nu}$. With the bulk background obeying the Einstein's field equations, the linearized equations Equ.~(\ref{eq:linearizedEF}) is invariant under the bulk diffeomorphism transform:
\begin{equation}
    h_{\mu\nu}(x,\rho)\rightarrow h_{\mu\nu}(x,\rho)+\nabla_{\mu}\epsilon_{\nu}(x,\rho)+\nabla_{\nu}\epsilon_{\mu}(x,\rho)\,.\label{eq:diffeo}
\end{equation}
We want to decompose the linearize bulk Einstein's equation Equ.~(\ref{eq:linearizedEF}) in to a set of equations with a clear 2d interpretation. This can be done by first defining the vector field $V_{\mu}(x,\rho)$ whose components are
\begin{equation}
    \begin{split}
   V_{\rho}(x,\rho)&=\frac{1}{2}\int_{\rho}^{\infty} du h_{\rho\rho}(x,u)\,, \\ V_{i}(x,\rho)&=e^{2A(\rho)}\int_{\rho}^{\infty} du e^{-2A(u)}h_{\rho i}(u,x)+\frac{e^{2A(\rho)}}{2}\int _{\rho}^{\infty} due^{-2A(u)}\int_{u}^{\infty} du' \partial_{i}h_{\rho\rho}(x,u')\,.\label{eq:Vdef}
    \end{split}
\end{equation}
This vector field is in fact the gravitational Wilson line in the $\rho$-direction and it is designed such that we have
\begin{equation}
    h_{\rho\mu}(x,\rho)=-\nabla_{\rho}V_{\mu}(x,\rho)-\nabla_{\mu}V_{\rho}(x,\rho)\,.
\end{equation}
This vector field transforms under the gauge transform Equ.~(\ref{eq:diffeo}) as
\begin{equation}
    V_{\mu}(x,\rho)\rightarrow V_{\mu}(x,\rho)-\epsilon_{\mu}(x,\rho)\,.\label{eq:Vtrans}
\end{equation}
Now we can use this vector field to define
\begin{equation}
    \bar{h}_{ij}(x,\rho)= h_{ij}(x,\rho)+2e^{2A(\rho)}A'(\rho)\bar{g}_{ij}(x)V_{\rho}(x,\rho)\,.\label{eq:newh}
\end{equation}
This new field is invariant under the transform Equ.~(\ref{eq:diffeo}) generated by the $\rho$-direction deformation $\epsilon_{\rho}(x,\rho)$ and it transforms under the $i$-direction diffeomorphisms as
\begin{equation}
    \bar{h}_{ij}(x,\rho)\rightarrow \bar{h}_{ij}(x,\rho)+\bar{\nabla}_{i}\epsilon_{j}(x,\rho)+\bar{\nabla}_{j}\epsilon_{i}(x,\rho)\,.
\end{equation}
Thus, this new field is more like a ``2d graviton" field than the old field. Then the bulk linearized equations can be decomposed as 
\begin{equation}
    \begin{split}
        0=&\frac{1}{2}e^{2A(\rho)}\partial_{\rho}\Big[e^{2A(\rho)}\partial_{\rho}e^{-2A(\rho)}(\bar{h}_{ij}-\bar{g}_{ij}\bar{h})\Big]+\frac{1}{2}e^{2A(\rho)}\partial_{\rho}\Big[e^{2A(\rho)}\partial_{\rho}e^{-2A(\rho)}(\bar{\nabla}_{i}V_{j}+\bar{\nabla}_{j}V_{i}-\bar{g}_{ij}2\bar{\nabla}_{i}V^{i})\Big]\,,\\0=&
        \frac{1}{2}e^{2A(\rho)}\partial_{\rho}\Big[e^{2A(\rho)}\partial_{\rho}e^{-2A(\rho)}\big[\bar{\nabla}^{j}\big(\bar{h}_{ij}+\bar{\nabla}_{i}V_{j}+\bar{\nabla}_{j}V_{i}\big)-\bar{\nabla}_{i}\big(\bar{h}+2\bar{\nabla}_{j}V^{j}\big)\big]\Big]\,,\\0=&\frac{1}{2}e^{2A(\rho)}\partial_{\rho}\Big[e^{2A(\rho)}\partial_{\rho}e^{-2A(\rho)}(\bar{h}+2\bar{\nabla}_{i}V^{i})\Big]\,,\label{eq:branegravitoneomsimplify}
    \end{split}
\end{equation}
where the upper indices are all lifted from lower indices using instead the 2d metric $\bar{g}^{ij}$. The fields $\bar{h}_{ij}(x,\rho)$ and $V_{i}(x,\rho)$ and the gauge parameter $\epsilon_{i}(x,\rho)$ obey the boundary conditions near the brane
\begin{equation}
    \partial_{\rho}e^{-2A(\rho)}\bar{h}_{ij}(x,\rho)_{\rho=\rho_{B}}=0\,,\quad\partial_{\rho}e^{-2A(\rho)}V_{i}(x,\rho)|_{\rho=\rho_{B}}=0\,,\quad\partial_{\rho}e^{-2A(\rho)}\epsilon_{i}(x,\rho)|_{\rho=\rho_{B}}=0\,.
\end{equation}
Thus, we can expand both of them using a set of Kaluza-Klein (KK) mode functions as the eigen-solutions of the following differential equation and boundary condition
\begin{equation}
    -e^{2A(\rho)}\partial_{\rho}e^{2A(\rho)}\partial_{\rho}e^{-2A(\rho)} \phi_{n}(\rho)=m_{n}^{2}\phi_{n}(\rho)\,,\quad\partial_{\rho}e^{-2A(\rho)}\phi_{n}(\rho)|_{\rho=\rho_{B}}=0\,.\label{eq:KKeomgraviton}
\end{equation}
From this differential equation and the boundary condition, the orthognality condition of the wavefunctions $\phi_{n}(\rho)$ is
\begin{equation}
    \int_{\rho_{B}}^{\infty} d\rho e^{-4A(\rho)}\phi_{n}(\rho)\phi_{m}(\rho)=\delta_{nm}\,,\label{eq:normgraviton}
\end{equation}
where we also normalized the wavefunctions. Hence, we can see that the zero mass eigenmode $\phi_{0}(\rho)=e^{2A(\rho)}=\cosh^{2}\rho$ is not normalizable. Projecting all the fields to the $r$-th KK mode, we have
\begin{equation}
    \begin{split}
       -\frac{m_{r}^{2}}{2}(\bar{h}_{ij}^{(r)}-\bar{g}_{ij}\bar{h}^{(r)})-\frac{m_{r}^{2}}{2}(\bar{\nabla}_{i}V_{j}^{(r)}+\bar{\nabla}_{j}V_{i}^{(r)}-\bar{g}_{ij}2\bar{\nabla}_{i}V^{(r)i})&=0\,,\\
    \frac{m_{r}^{2}}{2}\Big[\bar{\nabla}^{j}\big(\bar{h}^{(r)}_{ij}+\bar{\nabla}_{i}V^{(r)}_{j}\bar{\nabla}_{j}V^{(r)}_{i}\big)-\bar{\nabla}_{i}\big(\bar{h}^{(r)}+2\bar{\nabla}_{j}V^{(r)j}\big)\Big]&=0\,,\\
 m_{r}^{2}[\bar{h}^{(r)}+2\bar{\nabla}_{i}V^{(r)i}]&=0\,, 
 \end{split}
\end{equation}
where $m_{r}^{2}$ is the mass square of the $r$-th KK mode and the fields $\bar{h}^{(r)}$ and $V^{(r)i}$ are only functions of $x^{i}$, i.e. they are 2-dimensional fields. Defining $\tilde{h}^{(r)}_{ij}$ as
\begin{equation}
    \tilde{h}^{(r)}_{ij}=\bar{h}^{(r)}_{ij}+\bar{\nabla}_{i}V^{(r)}_{j}+\bar{\nabla}_{j}V^{(r)}_{i}\,,
\end{equation}
we have
\begin{equation}
       (\tilde{h}_{ij}^{(r)}-\bar{g}_{ij}\tilde{h}^{(r)})=0\,,\quad\bar{\nabla}^{j}\tilde{h}^{(r)}_{ij}-\bar{\nabla}_{i}\tilde{h}^{(r)}=0\,,\quad\tilde{h}^{(r)}=0\,, 
\end{equation}
for which the last equation is redundant with the first equation. These equations are exactly equations of motion of the St\"{u}ckelberg fields from the St\"{u}ckelberg mass term. Thus, we conclude that the brane localized gravity is purely a towel of modes with only S\"{u}ckelberg mass term
and no kinetic term. The absence of the kinetic term is not surprising due to the low-dimensional nature of the case we are considering. Moreover, the St\"{u}ckelberg/Goldstone vector fields are normalizable KK modes of the extra-dimensional gravitational Wilson line \cite{Geng:2025rov}.

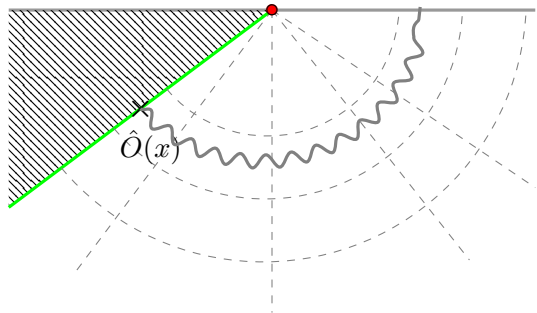
\begin{figure}
\begin{centering}
\begin{tikzpicture}[scale=1.4]
\draw[-,very thick,black!40] (-2.5,0) to (0,0);
\draw[-,very thick,black!40] (0,0) to (2.5,0);
\draw[pattern=north west lines,pattern color=black!200,draw=none] (0,0) to (-2.5,-1.875) to (-2.5,0) to (0,0);
%violet!50
\draw[-,dashed,color=black!50] (0,0) to (-1.875,-2.5); 
\draw[-,dashed,color=black!50] (0,0) to (0,-2.875);
\draw[-,dashed,color=black!50] (0,0) to (1.875,-2.375); 
\draw[-,dashed,color=black!50] (0,0) to (2.5,-1.6875); 
\draw[-,very thick,color=green!!50] (0,0) to (-2.5,-1.875);
%\draw[-] (-0.75,0) arc (180:217.5:0.75);
%\node at (-0.5,0.2) {$\mu$};
%\draw[-] (1.5,0) arc (0:-5.25:1.5);
\node at (0,0) {\textcolor{red}{$\bullet$}};
\node at (0,0) {\textcolor{black}{$\circ$}};
\draw[-,dashed,color=black!50] (-2,-1.5) arc (-140:-2:2.5);
\draw[-,dashed,color=black!50] (-1.5,-1.125) arc (-140:-2:1.875);
\draw[-,dashed,color=black!50] (-1,-0.75) arc (-140:-2:1.25);
\node at (-1.25,-0.9375) {\textcolor{black}{$\cross$}};
\node at (-1.15,-1.3) {\textcolor{black}{$\hat{O}(x)$}};
\draw[-,very thick,snake it,color=black!50] (-1.25,-0.9375) arc (-135:-5:1.56);
\end{tikzpicture}
\caption{A demonstration of the gravitational dressing from the bulk point of view. The waved curve is the gravitational Wilson line dressing the operator $\hat{O}(x)$ inserted on the brane to the bulk asymptotic boundary, i.e. the bath.}
\label{pic:branethetadress}
\end{centering}
\end{figure}

If one turns on other matter fields on the brane, which for example only couples to the $r$-th KK mode of the $\tilde{h}_{ij}$, then one would have
\begin{equation}
\begin{split}    
H_{\text{matter}}&=\frac{m_{r}^2}{16\pi G_{r}}\int dx (\bar{h}_{xx}+2\sqrt{16\pi G_{r}}\nabla_{x}V_{x})\,,\label{eq:massivegauss}
\end{split}
\end{equation}
where $G_{r}$ is the coupling strength of the matter with the $r$-th mode. Therefore, for the same reason as before, one is able to localize nontrivial information inside the AdS$_{2}$ brane from being detected on its asymptotic boundary. Nevertheless, from the AdS$_{3}$ perspective, this fact has a nice geometric origin. Firstly, the fact that the zero mass mode is not normalizable is important for this argument. As otherwise, we are forced to have no energetic excitations on the brane that are minimally coupled to the metric and so largely no nontrivial information at all. The fact that the zero mode is not normalizable is due to the extra dimensional geometry. This is because of the leftover asymptotic boundary of the AdS$_{3}$ bulk sits at infinity making the extra-dimension to the brane noncompact. This renders the zero mode nonnormalizable. Secondly, from the bulk point of view, we have the standard massless gravity so the gravitational Gauss' law still applies, i.e. one is able to detect the brane localized excitation from the asymptotic boundary of the bulk, i.e. the bath in the island model. The gravitational Gauss' law works geometrically as operators have to be dressed to obey the local diffeomorphism constraints, as we will review in some detail in Sec.~\ref{sec:necc}. The gravitational Wilson line connects the brane localized excitation to a point on the bulk asymptotic boundary and the energy is encoded at the ending point of the Wilson line on the asymptotic boundary (see Fig.~\ref{pic:branethetadress}). Thus, from the brane perspective, such an operator is not dressed to the asymptotic boundary of the brane, i.e. the defect, but rather to the bath. This is the reason why operators can be localized inside the island on the brane and are not able to be detected at the defect \cite{Geng:2025rov}.

\section{Massive Islands in General}\label{sec:in general}
In Sec.~\ref{sec:revisit}, we revisited the basics of massive islands using simple lower-dimensional models. We articulated how the graviton mass is extracted in the island model and the physical meaning of this mass together with its difference from the characteristic scale associated with energy dissipation. We noticed that the graviton mass allows information to be localized inside the gravitational AdS from being detected by the asymptotic boundary of the AdS. This point is essential for the consistency of the holographic interpretation of islands \cite{Geng:2021hlu,Geng:2023zhq,Geng:2025rov}. In this section, we will further clarify the above points in general contexts of the island model. We will see that the algorithm we applied in Sec.~\ref{sec:direct} to extract the graviton mass works in any dimensions \cite{Porrati:2001gx,Porrati:2002dt,Porrati:2003sa,Duff:2004wh,Aharony:2006hz,Geng:2023ynk,Geng:2023zhq,Geng:2025rov} and the essence of the graviton mass for the consistency of islands can be formulated more precisely \cite{Geng:2021hlu,Geng:2023zhq,Geng:2025rov}.

\subsection{Massive Islands in Higher Dimensions}\label{sec:higherdimensions}
The calculation of the graviton mass in higher-dimensional island models is the same as that we performed in Sec.~\ref{sec:direct}. Though the calculation is significantly more complicated than the lower-dimensional example we considered in Sec.~\ref{sec:direct}. Let's outline the calculation and the result in an example with a leaky free massive scalar field $\phi(x)$ minimally coupled to AdS$_{d+1}$ Einstein gravity.

The Green's function of this scalar field is given by
   \begin{equation}
\begin{split}
\langle \mathbf{T}\phi(x_{1})\phi(x_{2})\rangle&=a_{1}2^{-\Delta_{1}}\frac{\Gamma[\Delta_{1}]}{\pi^{\frac{d}{2}}(2\Delta_{1}-d)\Gamma[\Delta_{1}-\frac{d}{2}]}\frac{_{2}F_{1}[\frac{\Delta_{1}}{2},\frac{\Delta_{1}+1}{2},\Delta_{1}-\frac{d}{2}+1,\frac{1}{Z^{2}}]}{(Z)^{\Delta_{1}}}\\&+a_{2}2^{-\Delta_{2}}\frac{\Gamma[\Delta_{2}]}{\pi^{\frac{d}{2}}(2\Delta_{2}-d)\Gamma[\Delta_{2}-\frac{d}{2}]}\frac{_{2}F_{1}[\frac{\Delta_{2}}{2},\frac{\Delta_{2}+1}{2},\Delta_{2}-\frac{d}{2}+1,\frac{1}{Z^{2}}]}{(Z)^{\Delta_{2}}},
\end{split}
\end{equation}
where $Z$ is the invariant distance between the two operator insertions in AdS$_{d+1}$ and we have set the AdS$_{d+1}$ length scale to one with
\begin{equation}
    \Delta_{1}=\frac{d}{2}+\sqrt{\frac{d^2}{4}+m^2}\,,\quad \Delta_{2}=\frac{d}{2}-\sqrt{\frac{d^2}{4}+m^2}\,.
\end{equation}
The fact that this scalar is leaky is reflected by the fact that both $a_{1}$ and $a_{2}$ are nonzero. Following the same algorithm as Sec.~\ref{sec:direct}, one has to compute the large distance behavior of the stress-energy tensor two-point function of $\phi(x)$ and compare it with the Green's function of the massive vector field with mass squared $m^{2}_{V}=2d$. The complication comes from the requirement to decompose these two correlators into appropriate tensor structures. We can take the large distance limit and compare them only after the decomposition is done. Nevertheless, these decompositions were nicely done in \cite{Duff:2004wh,Aharony:2006hz}. Let's define
\begin{equation}
    \Pi^{\mu\nu,\rho\sigma}(x,y)=\langle \mathbf{T} T^{\mu\nu}(x) T^{\rho\sigma}(y)\rangle\,.
\end{equation}
It has been computed in \cite{Aharony:2006hz} that the large distance limit of this two point function is 
\begin{equation}
    \begin{split}
        16\pi G_{N}\Pi^{\mu\nu,\rho\sigma}(x,y)=-4M^{2}\nabla^{(\mu}G^{\nu),(\sigma}(x,y)\nabla^{\rho)}\,,\label{eq:TT}
        \end{split}
\end{equation}
where $G^{\nu,\sigma}(x,y)$ is in the form of the propagator of a massive vector field with mass squared $m^{2}_{V}=2d$. This vector field propagator is divergenceless and it satisfies
\begin{equation}
(\nabla^{2}-d)G^{\nu,\sigma}(x,y)=-i\frac{g^{0\nu\sigma}}{\sqrt{-g}}\delta^{d+1}(x-y)\,,\label{eq:Green}
\end{equation}
and $M^{2}$ is given in general in \cite{Aharony:2006hz} as
\begin{equation}
M^{2}=-G_{N}\frac{2^{4-d}\pi^{\frac{3-d}{2}}}{(d+2)\Gamma(\frac{d+3}{2})}\frac{a_{1}a_{2}\Delta_{1}\Delta_{2}\Gamma[\Delta_{1}]\Gamma[\Delta_{2}]}{\Gamma[\Delta_{1}-\frac{d}{2}]\Gamma[\Delta_{2}-\frac{d}{2}]}\,.\label{eq:mass}
\end{equation}
Then the same algorithm in Sec.~\ref{sec:direct} implies the induced graviton mass term
\begin{equation}
    S^{(2)}_{\text{eff},R}[h,V]=-\frac{M^2}{4}\int d^{2}x\sqrt{-g^0}\big(\tilde{h}_{\mu\nu}\tilde{h}^{\mu\nu}-\tilde{h}\tilde{h}\big)\,,\label{eq:massgeneral}
\end{equation}
which is in the Fierz-Pauli form and where we have the invariant combination of the graviton $h_{\mu\nu}(x)$ and the Goldstone vector field $V^{\mu}(x)$
\begin{equation}
    \tilde{h}_{\mu\nu}(x)=h_{\mu\nu}(x)+\nabla_{\mu}V_{\nu}(x)+\nabla_{\nu}V_{\mu}(x)\,.
\end{equation}
As a result, if we restore the dependence on the AdS$_{d+1}$ length scale $l_{\text{AdS}}$ and consider an arbitrary number $k$ of free leaky fields, the graviton mass scales as
\begin{equation}
    M^2 \sim \frac{k G_{N}}{l_{\text{AdS}}^{d+1}}\,.\label{eq:scaleofmass}
\end{equation}

\subsection{The Necessity of Massive Islands and Its Robustness}\label{sec:necc}
As we have discussed in Sec.~\ref{sec:CFTdissip}, the graviton mass enables information to be localized inside the gravitational AdS from being detected around its asymptotic boundary. In fact, this point is crucial for the consistency of the holographic interpretation of islands \cite{Geng:2021hlu,Geng:2023zhq,Geng:2025rov,Geng:2025tba1}. In this subsection, we will revisit and further articulate the necessity of graviton mass for islands.

To see the consistency conditions for the holographic interpretation of islands, let's consider the island model with an empty gravitational AdS$_{d+1}$ coupled to a $(d+1)$-dimensional nongravitational bath. Earlier studies have confirmed that islands do exist in this model \cite{Almheiri:2019yqk,Sully:2020pza,Geng:2020qvw,Geng:2021iyq} in various dimensions.

\subsubsection{The Necessity of Graviton Mass for Islands}\label{sec:review}
As we have seen, in the island model the graviton becomes massive due to the bath coupling. The emergent Goldstone vector boson $V^{\mu}(x)$ suggests that the mass is due to a Higgs mechanism with the AdS diffeomorphism symmetry spontaneously broken by the bath coupling. However, as opposed to the textbook treatment of Higgs mechanism, in our case we don't have a Higgs potential to start with and rather the Higgs mechanism in our case is a genuine quantum effect.\footnote{This is similar to the technicolor models \cite{Farhi:1980xs}.} However, the basic principle in the Higgs mechanism should carry over in our case. Namely, there should exist an order parameter which indicates the spontaneous diffeomorphism breaking in the AdS and the bulk diffeomorphism transform of this order parameter gives the Goldstone vector field $V^{\mu}(x)$. Let's firstly explain this order parameter and then see why it is crucial to see the necessity of graviton mass for islands.

As we have discussed, the spontaneous breaking of the AdS diffeomorphism should be due to the bath coupling. Moreover, the proper order parameter should have a nonzero vacuum expectation value (vev) that has a nontrivial dependence on the bulk positions in the AdS. This nontrivial AdS bulk position dependence indicates the spontaneous breaking of the AdS diffeomorphism symmetry. Thus, the natural order parameter to consider is a nonlocal operator
\begin{equation}
    \phi_{\text{AdS}}(x)\phi_{\text{bath}}(y)\,,\label{eq:trialorderparameter}
\end{equation}
where $\phi_{\text{AdS}}(x)$ is the leaky AdS field into a bath field $\phi_{\text{bath}}(y)$. The vev of this operator is nonzero due to the entanglement generated by the bath coupling and moreover this entanglement is strong enough due to the nonzero coupling such that this vev nontrivially depends nontrivially on the AdS position $x$ \cite{Geng:2025rov,Geng:2025yys}. In fact, this consideration suggests that there exist many such order parameters, as for example we can consider different bath positions $y$ and even smear the $y$ dependence with some kernels. In fact, a precise identification of this order parameter can only be done by properly applying the AdS/CFT correspondence (see \cite{Geng:2025rov}). Nevertheless, for our current purpose, the precise identification of such an order parameter is not important. The key point is that such order parameter is nonlocal involving local field in AdS and field in the bath. As a result, the Goldstone vector field $V^{\mu}(x)$ is also nonlocal which is however local from the perspective inside the AdS. This is because the Goldstone boson $V^{\mu}(x)$ is the diffeomorphism transform of the order parameter. For example, one can roughly think about Equ.~(\ref{eq:trialorderparameter}) as the order parameter and thus the Goldstone vector field
is in the following schematic form
\begin{equation}
    V^{\mu}(x)\sim \phi_{\text{bath}}(y)\partial^{\mu}\phi_{\text{AdS}}(x)\,.
\end{equation}
This point is crucial for the consistency of islands.

The operators inside the island should obey the diffeomorphism constraints which are the quantum version of the gravitational Gauss' law. The integrated Hamiltonian constraint can be shown to be of the form
\begin{equation}
    \sqrt{16\pi G_{N}}\Big(\hat{H}_{\text{matter}}+\hat{H}_{\text{graviton}}\Big)-\hat{\Pi}_{V^{0}}+\hat{H}_{\partial}=0\,,\label{eq:constraintMass}
\end{equation}
where $\hat{H}_{\partial}$ is a boundary term of the AdS metric fluctuation and is also called the ADM Hamiltonian \cite{Arnowitt:1962hi} and $\hat{\Pi}_{V^0}$ is the canonical conjugate of the 0-th component of the Goldstone vector field. The above equation should be obeyed as an operator equation. Thus, physical operators in AdS should commute with it, i.e.
\begin{equation}
    [\hat{O}^{\text{Phys}}(x),\sqrt{16\pi G_{N}}\Big(\hat{H}_{\text{matter}}+\hat{H}_{\text{graviton}}\Big)-\hat{\Pi}_{V^{0}}+\hat{H}_{\partial}]=0\,.
\end{equation}
In general, there are two ways to obey the above constraint
\begin{equation}
    [\hat{O}^{\text{Phys}}(x),\sqrt{16\pi G_{N}}\Big(\hat{H}_{\text{matter}}+\hat{H}_{\text{graviton}}\Big)]=-[\hat{O}^{\text{Phys}}(x),\hat{H}_{\partial}]\,,
\end{equation}
or
\begin{equation}
    [\hat{O}^{\text{Phys}}(x),\sqrt{16\pi G_{N}}\Big(\hat{H}_{\text{matter}}+\hat{H}_{\text{graviton}}\Big)]=[\hat{O}^{\text{Phys}}(x),\hat{\Pi}_{V^{0}}]\,.
\end{equation}
Using the fact that the matter Hamiltonian generates the time evolution of the matter fields, we have in the first case
\begin{equation}
    [\hat{O}^{\text{Phys}}(x),\hat{H}_{\partial}]=-i\sqrt{16\pi G_{N}}\partial_{t}\hat{O}^{\text{Phys}}(x)\,,
\end{equation}
and in the second case
\begin{equation}
    [\hat{O}^{\text{Phys}}(x),\hat{\Pi}_{V^0}]=i\sqrt{16\pi G_{N}}\partial_{t}\hat{O}^{\text{Phys}}(x)\,.
\end{equation}
In fact, in the first case the physical operator $\hat{O}^{\text{Phys}}(x)$ can be constructed using the gravitational Wilson line \cite{Donnelly:2016auv,Donnelly:2017jcd,Donnelly:2018nbv,Giddings:2018umg,Giddings:2019hjc} and in the second case it is easy to show that
\begin{equation}
    \hat{O}^{\text{Phys}}(x)=\hat{O}(x+\sqrt{16\pi G_{N}}\hat{V}(x))\,,\label{eq:islandoperator}
\end{equation}
where $\hat{O}(x)$ is a quantum field theory operator and $\hat{O}^{\text{Phys}}(x)$ is the gravitationally dressed operator. Operators inside the island should certainly fall into the second case, as in the first case the operator is then detectable using the boundary Hamiltonian from the AdS asymptotic boundary contradicting the factorization of the island from its AdS complement. The dressed operator in both the first and the second case can be shown to obey not only the Hamiltonian constraint Equ.~(\ref{eq:constraintMass}) but also the momentum constraints \cite{Geng:2025rov}.

Furthermore, it can be shown that the operator Equ.~(\ref{eq:islandoperator}) automatically obeys the second consistency condition for operators inside the island. The operator Equ.~(\ref{eq:islandoperator}) doesn't commute with the bath Hamiltonian. This is because, as we discussed, the Goldstone vector field $V^{\mu}(x)$ also has support on the bath, so it doesn't commute with the bath Hamiltonian. Thus, the energetic excitation created by the operator Equ.~(\ref{eq:islandoperator}) can be detected in the bath, which is another basic consistent condition for the holographic interpretation of islands. In fact, the operator Equ.~(\ref{eq:islandoperator}) can be interpreted as being dressed to the bath. This is especially clear if one thinks about the Karch-Randall braneworld. In the Karch-Randall braneworld, operators in the island live on the brane and they must be dressed using the Goldstone vector field $V^{\mu}(x)$ which is actually the extra-dimensional gravitational Wilson line in the dual bulk as we discussed in Sec.~\ref{sec:holographic}. The bulk description makes it clear that the operator Equ.~(\ref{eq:islandoperator}) is dressed to the bath through the extra dimension (see Fig.~\ref{pic:branethetadress}).

\subsubsection{Quantum Wormhole and the Graviton Mass}\label{sec:wormholemass}
Since we have reached an important point at the end of Sec.~\ref{sec:review} and some relevant comments exist in the literature, let's make some remarks.

As opposed to the argument at footnote 14 on page 15 of \cite{Engelhardt:2023xer}, we have explicitly shown that the coupling induced entanglement between the AdS and the bath is crucial for the consistency of islands. Importantly, such entanglement is strong enough to change the local physics in the gravitational AdS. The change of the local physics is indicated by the non-local order parameter, which in fact describes the local response of the AdS to an excitation of the bath. Even though the entanglement is nonlocal, this response extracted from the entanglement is causal and it is enabled by the coupling between AdS and bath. In other words, we have:
\begin{displayquote}
\textit{\textbf{With the bath decoupled from the AdS, one wouldn't have the same entanglement pattern as an implication of this entanglement pattern, i.e. causal response, is no longer satisfied.}} 
\end{displayquote}
Thus, the entanglement is strong enough to change the local physics inside AdS if and only if the AdS and the bath are coupled. More about the coupling/decoupling issue will be discussed in Sec.~\ref{sec:decoup}.

An interesting picture is obtained at the end of Sec.~\ref{sec:review} while we were thinking about the Goldstone vector field dressed operators in the Karch-Randall braneworld. In this case, there is a dual bulk description with a classical bulk geometry where the Goldstone vector field is identified as the gravitational Wilson line going through the extra dimension of the Karch-Randall brane. This suggests that in the island models without a classical bulk dual, the Goldstone vector field can be thought of as a \textit{quantum wormhole operator} that dresses operators inside the island to the bath through quantum extra dimensions. From this perspective, the Karch-Randall braneworld is the case where these quantum wormholes condense, which results in the classical bulk dual, and this is a manifestation of the emergence of the geometry from quantum entanglement \cite{Geng:2025rov}. In fact, this type of quantum wormhole operators constructed from quantum entanglement are ubiquitous \cite{Geng:2025tba1}.

\subsubsection{Robustness of the Quantum Wormhole}\label{sec:robust}
While a nice picture about the quantum wormhole operators as a result of quantum entanglement is obtained in \cite{Geng:2025rov}, there exist arguments in \cite{Antonini:2025sur} saying that the quantum entanglement that the quantum wormhole operator relies on is too subtle to be distillable and so the quantum wormhole operators are not generally physical.\footnote{See discussions in Sec. 4.7 in \cite{Antonini:2025sur}.} We will first try to formulate the arguments in \cite{Antonini:2025sur} in a precise way and then provide the resolution. A detailed example of the resolution will be presented in \cite{Geng:2025tba1}.

The essential point of \cite{Antonini:2025sur} is that order parameters like Equ.~(\ref{eq:trialorderparameter}) are not physically reasonable. This is because even though it has a nonzero vev with nontrivial dependence on the AdS position, it also has a large fluctuation around this vev. Thus, as opposed to the textbook Higgs mechanism where an order parameter has a sharply localized distribution around its symmetry breaking vacuum, typically Gaussian as the ground state wavefunction of a harmonic oscillator, it is not so clear if the naive order parameter Equ.~(\ref{eq:trialorderparameter}) really indicates the spontaneous breaking of the AdS diffeomorphism. To see the argument clearly, let's consider the example in Sec.~\ref{sec:higherdimensions}. In that case we could have free field $\phi_{\text{AdS}}(x)$ in the AdS leaking to a free field $\phi_{\text{bath}}(y)$ in the bath. Since the fields are free, tree-level propagators determine all correlation functions with Wick contractions. The tree-level propagators are determined by the transparent boundary condition between the AdS and the bath. Thus, we can compute the variance of the naive order parameter Equ.~(\ref{eq:trialorderparameter}) as
\begin{equation}
    \begin{split}
        &\langle\big(\phi_{\text{AdS}}(x)\phi_{\text{bath}}(y)\big)^2\rangle-\langle\phi_{\text{AdS}}(x)\phi_{\text{bath}}(y)\rangle^2\\&=\langle\phi_{\text{AdS}}(x)\phi_{\text{AdS}}(x)\rangle\langle\phi_{\text{bath}}(y)\phi_{\text{bath}}(y)\rangle+\langle\phi_{\text{AdS}}(x)\phi_{\text{bath}}(y)\rangle^2\,,\label{eq:variance}
    \end{split}
\end{equation}
which is much larger than the vev squared $\langle\phi_{\text{AdS}}(x)\phi_{\text{bath}}(y)\rangle^2$ as the first term on the right-hand-side is divergent due to the universal short-distance singularity of QFT correlators. Hence, one might conclude from here that the quantum wormhole operator, which is the fluctuation of the order parameter away from its vev in the direction generated by the AdS diffeomorphisms, is not a physically sensible operator.

However, as we discussed Equ.~(\ref{eq:trialorderparameter}) is not exactly the order parameter. In fact, one can consider properly smeared versions of Equ.~(\ref{eq:trialorderparameter}) which would have a small variation around its nontrivially AdS position dependent vev. This is rather easy to see physically as the large variance in Equ.~(\ref{eq:variance}) is due to the fact that infinitely many modes contribute to the real space correlators. Though in the actual measurements, we can only capture a few modes. Thus, the real order parameter is a properly smeared version of Equ.~(\ref{eq:trialorderparameter}). We refer readers to \cite{Geng:2025tba1} for an explicit construction of such a smeared operator.\footnote{We thank Daniel Jafferis and Neeraj Tata the intensive discussions about this point.}

Hence, we conclude that the quantum wormhole operators are robust and the quantum entanglement that generates the quantum wormhole operators is practically distillable. In fact, the above discussion resonates with our previous observation that the Higgs mechanism in the island model is a genuine quantum effect.

\section{The Physical Meaning of Graviton Mass}\label{sec:meaningmass}
As we have seen from the previous discussions and \cite{Geng:2025rov}, the graviton mass is generated by the Higgs mechanism associated with the spontaneous breaking of the diffeomorphism symmetry in AdS. Thus, with local experiments inside the AdS performed, one can confirm this Higgs mechanism by performing scattering experiments. These scatterings should be mediated by graviton. One can confirm the Higgs mechanism by comparing the scattering amplitudes in different sectors when the center of mass energy is above and below the characteristic energy scale of the spontaneous symmetry breaking. In AdS$_{4}$, below this scale there are five graviton polarization modes and the spin-0 mode dominates the scattering (assuming the center of mass energy is higher than the graviton mass) but above this scale there are only two graviton polarization modes and this spin-0 mode becomes decoupled. This is not surprising if one makes the analogy with the standard knowledge of the electroweak symmetry breaking in the Standard Model of particle physics. Nevertheless, this consideration suggests that the Higgs mechanism is a large distance effect and so should be the spontaneous symmetry breaking. This consideration brings in a puzzle whether one is able to distinguish the nonzero graviton mass from a massless graviton inside the AdS without going all the way to its asymptotic boundary. This is the question we will address in this section.

Let's first make a few quick remarks before we start. We note that holography itself is a long-range effect, as is the Higgs mechanism. Thus, one has to be careful about long-range vs short-range in gravity. The Higgs mechanism certainly modifies how holography works in the long-range sense. Strictly speaking, if one wants to understand holography one shouldn't talk about local short-range effects as such effects should eventually be completed to some long-range effects in quantum gravity. Though there is nothing wrong to expect more if one only cares about perturbative effects in gravity. Hence, in this section we focus on the short-range effects of the Higgs mechanism we uncovered and we will see that the Higgs mechanism indeed has nontrivial local signatures.

We first estimate the characteristic energy scale associated with the spontaneous diffeomorphism breaking. We will see that this energy scale is sufficiently bigger than both the graviton mass the inverse of the AdS length when one has a sizable island. Then we demonstrate that even in the usual Higgs mechanism in QFT, the boundary flux is not zero when the symmetry is spontaneously broken. These results are important to address various questions brought up in Section 2.3 of \cite{Antonini:2025sur}.

%Thus, the graviton mass is physical and is a genuine quantum effect as the standard Higgs mechanism in QFT. Hence, the graviton mass is different from the simple shift of the energy spectrum of a particle in a box when one changes the boundary condition of the particle near the box's boundary as argued in Section 2.3 of \cite{Antonini:2025sur}.

\subsection{The Energy Scale of Diffeomorphism Restoration}

\subsubsection{Estimating the Energy Scale}\label{sec:diffeobrokenscale}
As we have seen that in the island model, the graviton has an induced Feirz-Pauli mass term
\begin{equation}
    S_{\text{FP}}[h,V]=-\frac{M^2}{4}\int d^{d+1}x\sqrt{-g^{0}}\big(\tilde{h}_{\mu\nu}\tilde{h}^{\mu\nu}-\tilde{h}\tilde{h}\Big)\,,\label{eq:VFP}
\end{equation}
where $g^{0}_{\mu\nu}$ is the background metric which obeys vacuum Einstein's equation in AdS$_{d+1}$ and in terms of the linearized graviton field $h_{\mu\nu}(x)$ and the Goldstone vector field $V^{\mu}(x)$ we have
\begin{equation}
    \tilde{h}_{\mu\nu}(x)=h_{\mu\nu}(x)+\nabla_{\mu}V_{\nu}(x)+\nabla_{\nu}V_{\mu}(x)\,.
\end{equation}
We note that, from the graviton perspective, the Goldstone vector field looks to be a diffeomorphism transform and so the graviton kinetic term from linearizing the Einstein-Hilbert action is invariant if one replaces $h_{\mu\nu}(x)$ by $\tilde{h}_{\mu\nu}(x)$. Thus, at the quadratic level, the only $V^{\mu}(x)$-relevant action is Equ.~(\ref{eq:VFP}). Let's decompose the Goldstone vector field as
\begin{equation}
    V_{\mu}(x)=A_{\mu}(x)+\nabla_{\mu}\phi(x)\,,
\end{equation}
into transverse and longitudinal modes $A_{\mu}(x)$ and $\phi(x)$ \cite{Arkani-Hamed:2002bjr}. Let's focus on the longitudinal mode $\phi(x)$. The effective action Equ.~(\ref{eq:VFP}) can be expanded as
\begin{equation}
    \begin{split}
S_{\text{FP}}[h,V]&=-\frac{M^2}{4}\int d^{d+1}x\sqrt{-g}\Big(h_{\mu\nu}h^{\mu\nu}-h^2+4\nabla_{\mu}\nabla_{\nu}\phi\nabla^{\mu}\nabla^{\nu}\phi-4\Box\phi\Box\phi-4h\Box\phi+4h_{\mu\nu}\nabla^{\mu}\nabla^{\nu}\phi\Big)\,,\\&=-\frac{M^2}{4}\int d^{d+1}x\sqrt{-g}\Big(h_{\mu\nu}h^{\mu\nu}-h^2+4\frac{d(d+1)}{l_{AdS}^{2}}\nabla_{\mu}\phi(x)\nabla^{\mu}\phi(x)-4h\Box\phi+4\nabla^{\mu}\nabla^{\nu}h_{\mu\nu}\phi\Big)\,.\label{eq:kinetic}
    \end{split}
\end{equation}
Thus, we can see that the longitudinal modes has a kinetic term with coefficient $\frac{M^2 d(d+1)}{l^{2}_{AdS}}$ and it also mixes with the graviton field. Since we have $M\ll\frac{1}{l_{AdS}}$, the effect of the kinetic mixing with $h_{\mu\nu}$ to the kinetic term of the longitudinal mode $\phi(x)$ can be ignored.\footnote{This is because the graviton field has been canonical normalized so if one diagonalizes the graviton and $\phi(x)$ mixed kinetic term one will get a coefficient $M^{4}$ for the resulting $\phi(x)$ kinetic term. This term can be ignored comparing to the kinetic term of $\phi(x)$ we have in Equ.~(\ref{eq:kinetic}), due to $M^{2}\ll\frac{1}{l_{AdS}^2}$.} As a result, the canonically normalized longitudinal mode is
\begin{equation}
    \phi^{c}(x)=\frac{\sqrt{2d(d+1)}M}{l_{AdS}}\phi(x)\,.
\end{equation}
The interactions starts from terms of the form 
\begin{equation}
    S_{\text{int}}\sim M^2\sqrt{G_{N}}\int d^{d+1}x\sqrt{-g^0}(\nabla \nabla\phi)^3\sim\int d^{d+1}x\sqrt{-g^0}\sqrt{G_{N}}\frac{l^3_{AdS}}{M}(\nabla\nabla \phi^c)^3\,,\label{eq:interaction}
\end{equation}
which is due to the fact that at the next order in $G_{N}$ we have \cite{Arkani-Hamed:2002bjr}
\begin{equation}
    \tilde{h}_{\mu\nu}(x)=h_{\mu\nu}(x)+\nabla_{\mu}V_{\nu}(x)+\nabla_{\nu}V_{\mu}(x)+\sqrt{16\pi G_{N}}\nabla_{\mu}V^{\alpha}(x)\nabla_{\nu}V_{\alpha}(x)\,.
\end{equation}
Hence, the interaction becomes strong at the scale
\begin{equation}
    \Lambda_{\text{res}}\sim \Big(\frac{M}{l_{AdS}^3\sqrt{G_{N}}}\Big)^{\frac{2}{7+d}}\sim\Big(\frac{MM_{pl}^{\frac{d-1}{2}}}{l_{AdS}^3}\Big)^{\frac{2}{7+d}}\,,
\end{equation}
where $M_{pl}$ is the Planck scale. This is the diffeomorphism restoration scale, where the Goldstone vectors are no longer valid degrees of freedom \cite{Arkani-Hamed:2002bjr,Hinterbichler:2011tt}. Using our results about the graviton mass
\begin{equation}
    M\sim \frac{\sqrt{cG_{N}}}{l_{AdS}^{\frac{d+1}{2}}}\sim\frac{\sqrt{c}}{l_{AdS}}\Big(\frac{l_{pl}}{l_{AdS}}\Big)^{\frac{d-1}{2}}\,,
\end{equation}
we have
\begin{equation}
    \Lambda_{\text{res}}\sim\frac{c^{\frac{1}{7+d}}}{l_{AdS}}\,,
\end{equation}
where $c$ is the number of transparent fields.

\subsubsection{Implications}\label{sec:scattering}

In the semiclassical limit with a sizable island we have
\begin{equation}
    \frac{l_{pl}}{l_{AdS}}\ll1\,,\quad 1\ll c\ll\frac{l_{AdS}}{l_{pl}}\,.\label{eq:semiclassicallimit}
\end{equation}
As a result, in the above physically relevant limit we have
\begin{equation}
    M\ll \frac{1}{l_{AdS}}\ll\Lambda_{\text{res}}\ll\frac{1}{l_{pl}}\,.
\end{equation}
Thus, the graviton mass is physical even though its exact value might be small to be measurable by a local observer in AdS.
The essential point is that the correct scale the local observer should probe is $\Lambda_{\text{res}}$. This scale is large compared to the AdS localization scale $\frac{1}{l_{AdS}}$ when there is a sizable island. Thus, a local observer in AdS can confirm the Higgs mechanism by doing scattering experiments of the graviton and studying the amplitude of the spin-0 channel. The sharp signature of the Higgs mechanism in this scattering channel is that the scattering amplitude in this channel initially grows as the center of mass energy is dialed up, due to the derivative coupling we see in the low energy effective action Equ.~(\ref{eq:interaction}), but it starts decreasing when the center of mass energy is approaching the scale $\Lambda_{\text{res}}$ due to the Higgs cancellation effect \cite{Lee:1977eg,Lee:1977yc}.

We note that probing the Higgs mechanism through such scatterings is in fact very cheap. This is because the diffeomorphism restoration scale $\Lambda_{\text{res}}$ is much less than the Planck scale $M_{pl}=\frac{1}{l_{pl}}$.

\subsection{More on the Local Distinguishability between Massive and Massless Graviton}\label{sec:massislocal}

We note that a typical argument against the graviton mass as a meaningfully local concept in the island model is the following:\footnote{See for example the discussions on the bottom of page 14 in \cite{Antonini:2025sur}.}
\begin{displayquote}
\textit{The value of the graviton mass is hard to measure locally due to $Ml_{AdS}\ll1$ but it can be extracted from measuring the near-boundary decaying behavior of the graviton's Green's function, thus the graviton mass is only a globally defined concept but not locally.}
\end{displayquote}
In fact, this type of argument is not self-consistent. The reason is that the indistinguishability of the graviton propagator from the massless one in the strict $M l_{AdS}\rightarrow0$ limit is independent of the probing scale. This is due to the absence of the VDVZ discontinuity in AdS \cite{Porrati:2000cp}.\footnote{In flat space, there is VDVZ discontinuity and one is able to distinguish massive and massless graviton even in the strict $M l_{AdS}\rightarrow0$ limit.} Thus, in this limit one cannot distinguish the graviton propagator from the massless one no matter how close one is to the AdS boundary. More precisely, in this limit the asymptotic fall-off behavior of the graviton's Green's function is also the same as the massless case. This is because the near boundary fall-off behavior of the graviton's Green's function is \cite{Kabat:2012hp}
\begin{equation}
    G_{ij,kl}(x,z;y,z')\sim z^{\Delta-2}\,,\quad\text{as $z\rightarrow0$}\,,\label{eq:gravitonfalloff}
\end{equation}
where we consider the AdS$_{d+1}$ Poincar\'{e} patch and the non-radial components of the graviton. The conformal weight $\Delta$ is given by\footnote{We note that this following formula is standard in studies of AdS/CFT. Though it is different from Equation (2.7) in \cite{Antonini:2025sur} as the authors of \cite{Antonini:2025sur} used a different convention to define the mass for spinning particles in AdS. That definition in \cite{Antonini:2025sur} is not consistent with the general covariance principle in the conventional form.}
\begin{equation}
    \Delta=\frac{d}{2}+\sqrt{\frac{d^2}{4}+M^2 l_{AdS}^2}\,.\label{eq:exponent}
\end{equation}
Hence, in the strict $Ml_{AdS}\rightarrow0$ limit, the asymptotic fall-off behavior of the graviton's Green's function Equ.~(\ref{eq:gravitonfalloff}) doesn't discriminate massive and massless gravitons.

Thus, the issue is really if one's measurement is able to resolve the small number $Ml_{AdS}$. If one is able to resolve this small number by measuring the difference between the asymptotic fall-off behavior of the graviton Green's function with the massless case, then one is equally well able to locally distinguish the difference between the graviton propagator from the massless case inside the AdS bulk. More explicitly, one can consider the case of AdS$_4$ for which the graviton propagator has been worked out in \cite{Porrati:2000cp} as
\begin{equation}
\begin{split}
&2T_{\mu\nu}'(\Delta_{L}^{(2)}+M^2-2\Lambda)^{-1}T^{\mu\nu}-\frac{2}{3}T'(-\nabla^2+M^2-2\Lambda)^{-1}T\\&+\frac{2\Lambda}{9}T'(-\nabla^2+M^2-2\Lambda)^{-1}(\nabla^2+\frac{4\Lambda}{3})^{-1}T+\frac{2}{6-9\frac{M^2}{\Lambda}}T'(\nabla^2+\frac{4\Lambda}{3})^{-1}T\,,\label{eq:gravitonprop}
\end{split}
\end{equation}
where $T_{\mu\nu}$ is a probe matter stress-energy tensor with trace $T$ under the background AdS metric, $\Delta_{L}^{(2)}$ is the Lichnerowicz operator acting on spin-2 symmetric tensors and $\Lambda=-\frac{3}{l_{AdS}^2}$ is the cosmological constant for AdS$_{4}$. Therefore, one can see that in the limit that the probe energy scale is high, i.e. $\Delta_{L}^{(2)},\nabla^2\gg\Lambda,M^2$, one has the graviton propagator
\begin{equation}
\begin{split}
&2T_{\mu\nu}'(\Delta_{L}^{(2)})^{-1}T^{\mu\nu}-\frac{2}{3}T'(-\nabla^2)^{-1}T+\frac{2}{6-9\frac{M^2}{\Lambda}}T'(\nabla^2)^{-1}T+\mathcal{O}(\frac{\Lambda}{\Delta_{L}^{(2)}},\frac{\Lambda}{\nabla^2},\frac{M^2}{\Delta_{L}^{(2)}},\frac{M^2}{\nabla^2})\,.\label{eq:localprobe}
\end{split}
\end{equation}
The difference between the high-energy graviton propagator Equ.~(\ref{eq:localprobe}) with the massless case is purely controlled by the last term, i.e.
\begin{equation}
    \frac{2}{6-9\frac{M^2}{\Lambda}}T'(\nabla^2)^{-1}T= \frac{2}{6+3M^2l_{AdS}^2}T'(\nabla^2)^{-1}T=(\frac{1}{3}-\frac{1}{6}M^2l_{AdS}^2+\cdots)T'(\nabla^2)^{-1}T.
\end{equation}
As a result, this difference is also controlled by the small number $M l_{AdS}$ and the leading difference is of order $M^2 l_{AdS}^2$ which is the same as the difference between the asymptotic fall-off exponent Equ.~(\ref{eq:exponent}) with the massless case. 

In summary, one is able to locally measure the value of the graviton mass inside AdS, as long as one is able to differentiate the asymptotic fall-off behavior of the graviton's Green's function from the massless case. However, we note that for the purpose to locally confirm the nonzero graviton mass, the scattering protocol in Sec.~\ref{sec:scattering} is cheaper than directly measuring the graviton mass as in this subsection. This is because the characteristic dimensionless number in the scattering protocol $\Lambda_{\text{res}}l_{AdS}$ is much bigger than $Ml_{AdS}$ in the physically relevant regime Equ.~(\ref{eq:semiclassicallimit}). Moreover, directly measuring the graviton mass as in this subsection is not super expensive when there is a sizable island as in this case $Ml_{AdS}\gg\frac{l_{pl}}{l_{AdS}}$, so one doesn't have be able to probe all the way to the Planck scale.

\subsection{The Boundary Flux in the Standard Higgs Mechanism}\label{sec:flux}
The argument we collected at the beginning of Sec.~\ref{sec:massislocal} is oftentimes accompanied by the following argument as its ``sanity check":

\begin{displayquote}
    \textit{In the island model, the graviton mass is due to the nonvanishing boundary flux. So it makes sense that this mass is only a near boundary effect. As one can measure this flux by going close to the AdS boundary and so we shouldn't be able to be more powerful than we are supposed to be to locally measure the nonzero graviton mass.}
\end{displayquote}
In fact, there are two issues with the above assertion. First, as we discussed in Sec.~\ref{sec:massvsdiss} the graviton mass is an equilibrium effect, so if one simply measures the expectation value of the flux, one gets zero, as we are in the equilibrium state. Thus, to measure this transparent boundary condition one must disturb the system and observe its response. For example, one, as an observer in the AdS, can create an excitation near the boundary and track where it goes. As an AdS observer, one has to go into the bulk of the AdS to see that this excitation didn't go into the bulk and conclude that it leaks out of the AdS. Thus, it is not so clear if one can reasonably ``measure the flux by going to the AdS boundary". Second, even in the standard textbook Higgs mechanism the boundary flux of the broken symmetry is not zero as an operator equation. For example, in the Abelian-Higgs model the U(1) current in the Higgs phase has a boundary flux density given by
\begin{equation}
    J^{r}=v^2 (\partial_{r}\theta(x)-eA_{r}(x))\,,
\end{equation}
where we use the spherical coordinate system for a flat space with $r$ the radial direction and $v$ is the symmetry breaking vev of the complex scalar field with $\theta(x)$ the U(1) Goldstone boson. When $r$ is large we can ignore $A_{r}(x)$ due to the Yukawa exponential decay. Thus we have\footnote{Only the zero angular momentum mode is relevant for the flux computation in Equ.~(\ref{eq:flux}).}
\begin{equation}
    J^{r}=v^2\partial_{r}\theta(x)\sim \frac{v^2}{r^{d-1}}\,,
\end{equation}
where we used the fact that $\theta(x)$ is massless and the flux is given by
\begin{equation}
    \int J^{r} r^{d-1}d\Omega_{d}\sim v^2\Omega_{d}\,,\label{eq:flux}
\end{equation}
which is a constant, exactly the same case as in the island model. Thus, if one bought the above argument, one would conclude that in the standard Higgs mechanism one would only be able to confirm the Higgs phase by going to infinity. However, as we know, this is not the case, so we addressed the above argument by contradiction.

\subsection{AdS vs a Box/Cavity and Graviton vs Photon}\label{sec:notjustabox}
 It is standard in the study of the AdS/CFT correspondence to think of the AdS as a box with size $l_{AdS}$ which regularizes the flat space in the IR. It might be tempting to take this analogy seriously and further argue that the graviton mass is due to the change of the boundary condition near the boundary of this box and so it should be of order $\frac{1}{l_{AdS}}$.\footnote{Similar argument was also made in the prime version of \cite{Antonini:2025sur}. Here we take the opportunity to clarify this point for future reference.} In fact, as we have noticed in Sec.~\ref{sec:direct} and Sec.~\ref{sec:higherdimensions} that the graviton mass is a gravitational effect due to the local gravitational interaction between matter and graviton. It is not simply $\frac{1}{l_{\text{AdS}}}$ but much smaller than it. However, as we have discussed in Sec.~\ref{sec:scattering} and Sec.~\ref{sec:massislocal}, this fact doesn't obstruct one from locally confirming this nonzero mass or even measuring its value. 

 Moreover, we should notice that the fact that the graviton has a nonzero mass in the island model has no analogy in cavity QED for a flat box with size L. One could think that a baby version of the graviton mass in the island model is the photon mass, i.e. one can consider a U(1) gauge boson with charged matter fields inside the AdS and the bath coupling enables the U(1) current to flow into the bath. Indeed, the photon obtains a nonzero mass in this scenario due to almost the same reason as the graviton mass \cite{Rattazzi:2009ux,Karch:2023wui,Geng:2025tba2}. However, one might further argue that this effect has an analogy in cavity QED as one could consider the cavity to be in contact with an outside environment such that the electron inside the cavity can freely leak outside and speculate that this environment coupling might induce a mass for the photon.\footnote{Similar argument was made in the prime version of \cite{Antonini:2025sur}. We also thank Henry Maxfield for discussions.} This speculation is reasonable, however the induced photon mass in cavity QED is of a totally different nature and it cannot be interpreted as a Higgs mechanism as in the island model. In the open cavity case we can consider the totally transparent boundary condition such that the charged matters are immune to the cavity. Thus, as opposed to the island model where one has bound states inside the AdS from the transparent matter modes, there is no such bound states for open cavity QED. The existence of these bound states is a particular feature of the representation theory of the AdS isometry group together with the transparent boundary condition. The AdS transparent boundary condition implies that there are two different irreducible representations associated with each transparent matter field. The bound states are components of the tensor products of these distinct irreducible representations and one of these components is the Goldstone boson associated with the spontaneous symmetry breaking \cite{Porrati:2003sa}. In the open cavity case, the charged matters are living on the flat Minkowski space and there is only one irreducible representation of the Poincar\'{e} group associated with each leaky charged matter and so there are no bound states. Thus, one doesn't have an analogous Higgs mechanism in the cavity QED as in the island model due to the absence of the Goldstone boson. 
 
 In fact, there is a sharp distinction between the photon and graviton in the island model. Even though the photon (or more generally the gluon) becomes massive due to the same Higgs mechanism, one may not be able to locally distinguish the difference between the massive photon propagator from the graviton propagator as in Sec.~\ref{sec:scattering} or Sec.~\ref{sec:massislocal}. First, the symmetry restoration scale for the gauge symmetry is
 \begin{equation}
     \Lambda_{\text{res}\gamma}=\frac{M_{\gamma}}{g_{e}}\sim\frac{\sqrt{c'}}{l_{AdS}}\,,
 \end{equation}
 where $g_{e}$ is the gauge coupling strength and $c'$ denotes the number of transparent charged matter fields weighted by their charge square. Thus, if one only has a few transparent charged matter fields, this symmetry restoration scale is of the same order as the inverse AdS length scale. Thus, in these cases, once one probes into short distance, the gauge symmetry is effectively restored and one is not able to confirm the Higgs mechanism by doing local scattering experiments. We leave the question of whether there is a good reason for $c'$ to be large to future research. Second, unlike the graviton, the photon doesn't have a trace mode and the transverse mode decouples from the conserved current so the analogous propagator to Equ.~(\ref{eq:gravitonprop}) to photon is 
 \begin{equation}
     J'_{\mu}(-\Delta_{L}^{(1)}+M_{\gamma}^2-2\Lambda)^{-1}J^{\mu}\,,
 \end{equation}
where $\Delta_{L}^{(1)}$ is the Lichnerowicz operator acting on vector fields. Thus, when the probe scale is very high the propagator just becomes the flat space massless photon propagator to all orders in $M_{\gamma} l_{AdS}$. Hence, in summary, it might be hard to locally distinguish the photon mass from being nonzero in the island model compared to the graviton. Physically, this is not surprising as the issue of locality, i.e. the existence of local observables, is sharper in gravity than in gauge theory.\footnote{This is because in gauge theory one can have decent local gauge invariant observables. For example, positive charged operator dressed to a nearby negative charged operator by a short Wilson line or the field strength (in non-Abelian case with a simple gauge group one has $\Tr F^n$ for $n\geq2$).}

\section{Comment on Attempts for Massless Islands}\label{sec:attempts}
In previous sections, we have revisited the massive islands story, clarified potential confusions, and addressed various questions and counter-arguments regarding massive islands that exist in and out of the literature. Nevertheless, we notice that there are various proposed examples of islands in massless gravity in the literature and less articulated questions and counter-arguments regarding massive islands. We will address these issues in this section by carefully analyzing those proposed examples, questions and counter-arguments. As we will see, the analysis in this section suggests many interesting questions to explore in the future.

\subsection{Coarse Graining}
The most straightforward reflection about the massive islands story might be that it is a fine-grained statement with gravitational effects carefully incorporated and what if one just ignores some gravitational effects, i.e. consider some coarse-grained questions. Various proposed examples of entanglement islands in massless gravity exist in the literature falling into this category intentionally \cite{Dong:2020uxp,Krishnan:2020fer,Ghosh:2021axl,Antonini:2025sur} or unintentionally \cite{Miao:2022kve,Miao:2023unv}.

The common feature of these examples is that with a careful application of the QES formula with gravitational effects carefully incorporated one can easily see that the island is empty \cite{Geng:2020fxl,Geng:2023qwm}. A potential argument for the consistency of these examples is that one could consider the algebra consisting of operators but with the ADM Hamiltonian discarded. Then one speculates that this algebra might have an island. However, one should notice that the QES formula could give an island only if one didn't take $G_{N}$ as strictly zero, otherwise the entanglement between $\mathcal{I}$ and $R$ can never compete with the $\frac{A(\partial\mathcal{I})}{4G_{N}}$ term in the QES formula Equ.~(\ref{eq:islandformula}), and so a nontrivial island cannot be produced. However, with $G_{N}$ not strictly zero the above set of operators doesn't form an algebra as they are not closed under multiplication. This is because the operator product expansion (OPE) of those operators will contain the ADM Hamiltonian at order $\mathcal{O}(G_{N})$ due to the universal coupling of the gravitational force.

We should mention that there is one exceptional case to the above consideration-- the future null infinity in asymptotically flat space \cite{Laddha:2020kvp}. In this case, one has a set of free operators and their canonical conjugates at null infinity. Thus, one might be able to consistently discard the ADM Hamiltonian, which in this context is the same as the Bondi mass at $u=-\infty$, from this asymptotic algebra.\footnote{This observation had been made in \cite{Laddha:2020kvp} and reviewed in \cite{Raju:2020smc} (see page 84-86). This point was also revisited recently in Section. 3.2 of \cite{Antonini:2025sur}.} Thus, one might hope to have a nontrivial Page curve for a subalgebra of this algebra which consists of these free operators in $(-\infty,u)$ along the future null infinity as a function of $u$. However, as long as one goes slightly into the bulk, this free field structure will be broken due to the gravitational interaction and the ADM Hamiltonian will appear in the OPE of other operators. Thus, the practical relevance of this particular exceptional case is not very clear.

Hence, it would be interesting to think more carefully about this class of examples in the future to see if there is any self-consistent coarse-graining protocol with practical relevance that produces nontrivial entanglement islands, and to show that gravitational effects kick in in a subtler way to reduce the size of the island to empty.

\subsection{Island as a Property of Quantum State}

An interesting intuition about entanglement islands is that they are properties of the quantum states, as islands are produced by the QES formula due to the large entanglement between $\mathcal{I}$ and $R$. Thus, one might conclude from here that entanglement islands have nothing to do with graviton mass as one might identify graviton mass as a property of the dynamics of the system because it is produced by coupling the AdS to a nongravitational bath. However, as we saw in Sec.~\ref{sec:in general} that graviton mass is from the spontaneous diffeomorphism breaking and it is well understood that spontaneous symmetry breaking is a property of the quantum state. Therefore, the above insight actually suggested a deep connection between entanglement island and graviton mass. In fact, it is also part of the motivation for the discovery of the quantum wormhole operator in \cite{Geng:2025rov} which is, as we discussed, due to the detailed entanglement structure between $\mathcal{I}$ and R.

\subsection{Decoupling the Bath}\label{sec:decoup}
A very interesting argument exists supporting the idea that entanglement islands are properties of the quantum states.\footnote{This argument is from Juan Maldacena and partly discussed in \cite{Almheiri:2019yqk}. We thank Juan Maldacena for pointing this out and thorough discussions on it.} We will revisit this argument and explain why it substantiates the massive islands story. 

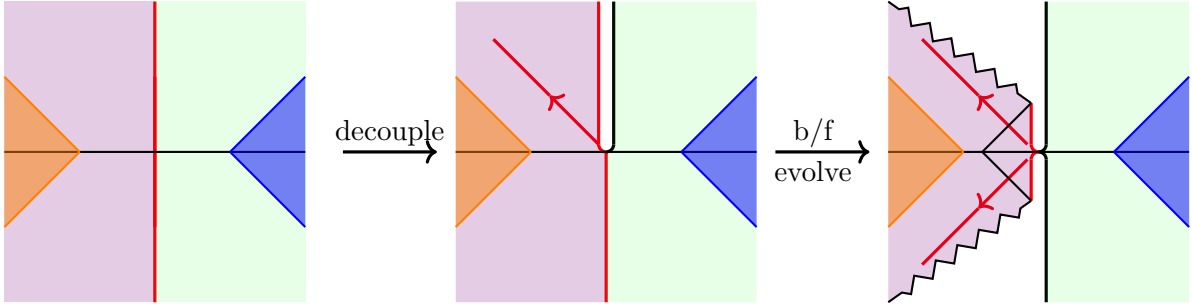
\begin{figure}[h]
    \centering
    \begin{tikzpicture}[scale=0.5]
       \draw[-,very thick,red](0-1,-4) to (0-1,4);
       \draw[fill=green, draw=none, fill opacity = 0.1] (0-1,-4) to (4-1,-4) to (4-1,4) to (0-1,4);
           \draw[-,very thick,red](0-1,-2) to (0-1,2);
       \draw[fill=violet, draw=none, fill opacity = 0.2] (0-1,-4) to (-4-1,-4) to (-4-1,4) to (0-1,4);
       \draw[-,thick,black] (-4-1,0) to (4-1,0);
       \draw[-,thick,blue] (2-1,0) to (4-1,2);
       \draw[-,thick,blue] (2-1,0) to (4-1,-2);
        \draw[fill=blue, draw=none, fill opacity = 0.5] (2-1,0) to (4-1,2) to (4-1,-2) to (2-1,0);
        \draw[-,thick,orange] (-2-1,0) to (-4-1,2);
       \draw[-,thick,orange] (-2-1,0) to (-4-1,-2);
        \draw[fill=orange, draw=none, fill opacity = 0.5] (-2-1,0) to (-4-1,2) to (-4-1,-2) to (-2-1,0);
        %%%%%%%%%%%%%%%%%%%%%%
        \draw[->,very thick,black] (4.5-0.5,0) to (6.5,0);
         \node at (5.25,0.5) {\textcolor{black}{decouple}};
        %%%%%%%%%%%%%%%%%%%%%%
        \draw[-,very thick,red](0+11,-4) to (0+11,0);
        \draw[-,very thick,red](-0.2+11,4) to (-0.2+11,0.2);
        \draw[-,very thick,black](0.2+11,4) to (0.2+11,0.2);
        \draw[->,very thick,red]  (-0.2+11,0.2) to (-1.5+11,1.5);
        \draw[-,very thick,red]  (-1.5+11,1.5) to (-3+11,3);
        \draw[-,very thick,red] (-0.2+11,0.2) arc (180:270:0.2);
        \draw[-,very thick,black] (0+11,0) arc (270:360:0.2);
       \draw[fill=green, draw=none, fill opacity = 0.1] (0+11,-4) to (4+11,-4) to (4+11,4) to (0.2+11,4) to (0.2+11,0.2) arc (0:-90:0.2) to (0+11,-4);
       \draw[fill=violet, draw=none, fill opacity = 0.2] (0+11,-4) to (-4+11,-4) to (-4+11,4) to (-0.2+11,4) to (-0.2+11,0.2) arc (180:270:0.2) to (0+11,-4);
       \draw[-,thick,black] (-4+11,0) to (4+11,0);
       \draw[-,thick,blue] (2+11,0) to (4+11,2);
       \draw[-,thick,blue] (2+11,0) to (4+11,-2);
        \draw[fill=blue, draw=none, fill opacity = 0.5] (2+11,0) to (4+11,2) to (4+11,-2) to (2+11,0);
        \draw[-,thick,orange] (-2+11,0) to (-4+11,2);
       \draw[-,thick,orange] (-2+11,0) to (-4+11,-2);
        \draw[fill=orange, draw=none, fill opacity = 0.5] (-2+11,0) to (-4+11,2) to (-4+11,-2) to (-2+11,0);
        %%%%%%%%%%%%%%%%%%%%%
        \draw[->,very thick,black] (4.5+11,0) to (6.5+11+0.5,0);
         \node at (5.5+11,0.5) {\textcolor{black}{b/f}};
         \node at (5.5+11,-0.5) {\textcolor{black}{evolve}};
        %%%%%%%%%%%%%%%%%%%%%%
         \draw[->,very thick,red]  (-0.2+22+0.4,0.2) to (-1.5+22+0.4,1.5);
        \draw[-,very thick,red]  (-1.5+22+0.4,1.5) to (-3+22+0.4,3);
        \draw[->,very thick,red]  (-0.2+22+0.4,-0.2) to (-1.5+22+0.4,-1.5);
        \draw[-,very thick,red]  (-1.5+22+0.4,-1.5) to (-3+22+0.4,-3);
        \draw[-,very thick,red](-0.2+22+0.5,-1.3) to (-0.2+22+0.5,-0.2);
        \draw[-,very thick,red](-0.2+22+0.5,1.3) to (-0.2+22+0.5,0.2);
        \draw[-,very thick,black](0.2+22+0.5,-4) to (0.2+22+0.5,-0.2);
        \draw[-,very thick,black](0.2+22+0.5,4) to (0.2+22+0.5,0.2);
        \draw[-,very thick,red] (-0.2+22+0.5,0.2) arc (180:270:0.2);
        \draw[-,very thick,red] (-0.2+22+0.5,-0.2) arc (180:90:0.2);
         \draw[-,very thick,black] (22+0.5,0) arc (270:360:0.2);
        \draw[-,very thick,black] (22+0.5,0) arc (90:0:0.2);
       \draw[fill=green, draw=none, fill opacity = 0.1] (0.2+22+0.5,-4) to (4+22+0.5,-4) to (4+22+0.5,4) to (0.2+22+0.5,4) to (0.2+22+0.5,0.2) arc (0:-90:0.2) arc (90:0:0.2) to (0.2+22+0.5,-4);
       \draw[fill=violet, draw=none, fill opacity = 0.2] decorate[decoration={zigzag,pre=lineto,pre length=5pt,post=lineto,post length=5pt}] {(-4+22+0.5,4) to (-1.5+22+0.5+1.5-0.2,1.3)}decorate[decoration={zigzag,pre=lineto,pre length=5pt,post=lineto,post length=5pt}] {(-1.5+22+0.5+1.5-0.2,-1.3) to (-4+22+0.5,-4)} (0+22+0.5-0.2,-1.3) to (-4+22+0.5,-4) to (-4+22+0.5,4) to (0+22+0.5-0.2,1.3) to (0+22+0.5-0.2,0.2) arc (180:270:0.2) arc (90:180:0.2) to (0+22+0.5-0.2,-4);
       \draw[-,thick,black] (-4+22+0.5,0) to (4+22+0.5,0);
       \draw[-,thick,blue] (2+22+0.5,0) to (4+22+0.5,2);
       \draw[-,thick,blue] (2+22+0.5,0) to (4+22+0.5,-2);
        \draw[fill=blue, draw=none, fill opacity = 0.5] (2+22+0.5,0) to (4+22+0.5,2) to (4+22+0.5,-2) to (2+22+0.5,0);
        \draw[-,thick,orange] (-2+22+0.5,0) to (-4+22+0.5,2);
       \draw[-,thick,orange] (-2+22+0.5,0) to (-4+22+0.5,-2);
       \draw[-,thick,black] (-1.5+22+0.5,0) to (-1.5+22+0.5+1.5-0.2,1.3);
       \draw[-,thick,black] (-1.5+22+0.5,0) to (-1.5+22+0.5+1.5-0.2,-1.3);
        \draw[fill=orange, draw=none, fill opacity = 0.5] (-2+22+0.5,0) to (-4+22+0.5,2) to (-4+22+0.5,-2) to (-2+22+0.5,0);
        \draw[-,thick] 
       decorate[decoration={zigzag,pre=lineto,pre length=5pt,post=lineto,post length=5pt}] {(-4+22+0.5,4) to (-1.5+22+0.5+1.5-0.2,1.3)};
       \draw[-, thick] 
       decorate[decoration={zigzag,pre=lineto,pre length=5pt,post=lineto,post length=5pt}] {(-1.5+22+0.5+1.5-0.2,-1.3) to (-4+22+0.5,-4)};
    \end{tikzpicture}
    \caption{\small{Let's consider the island model with a empty gravitational AdS (the purple region) coupled to a bath (the green region). Let's take a Cauchy slice (the black horizon slice) and consider the bath subregion $R$. The domain of dependence of the bath subregion is the blue shaded region. Its entanglement entropy is computed by Equ.~(\ref{eq:islandformula}). The orange region in the gravitational bulk denotes domain of dependence of the resulting island $\mathcal{I}$. Let's consider the following operations. We first decouple the AdS and the bath at the instant specified by the Cauchy slice we are considering. A shock wave is generated by due to the decoupling. Then we evolve the state on the Cauchy slice back using the decoupled Hamiltonian. The resulting Penrose diagram is provided on the right, where a black hole and white hole are formed due to the shockwave and island is behind the black hole horizon. The waved lines are black hole and white hole singularities. However, the local physics in between the two shockwaves are not changed due to causality.}}
    \label{pic:bathsubregion}
\end{figure}

The goal of this argument is to isolate the relevant physics of the island with irrelevant parts modded out. This is a more apparent way to resolve the confusion identifying the graviton mass as a property of the dynamics of the system induced by the bath but not the quantum state.\footnote{In fact, this confusion is not unreasonable as we do need to assume that the dynamics is relativistic.} One can isolate the quantum state in the scenario as depicted in Fig.~\ref{pic:bathsubregion}. We started from a state in the island model which is specified on a Cauchy slice through the union of the AdS and the bath on which there is an island due to Equ.~(\ref{eq:islandformula}). Then we can suddenly decouple the bath after the instant specified by this Cauchy slice. This decoupling switched the dynamics inside the AdS back to that without a bath and it induced a positive energy shockwave injected into the AdS from the decoupling point. The shockwave is lightlike due to the absence of a characteristic energy scale. Then we can evolve the state forward and backward using the decoupled Hamiltonian. The situation inside the AdS is depicted on the far right of Fig.~\ref{pic:bathsubregion}. A black hole and a white hole will be formed due to the shockwave, and the entanglement island now resides behind the double horizon. Nevertheless, due to causality, nothing changed in the spacelike region between the two shockwaves. For example, the matter field correlation functions with insertions only in this region or with this region and the bath are not changed at all and neither will the geometry be changed. This is also the region where the entanglement island lives. Thus, the same calculation of graviton mass carries over in this region with the same nonzero result. More succinctly, this decoupling doesn't change the entanglement structure between the island region and the bath. As we have discussed, this is the essential reason both for the existence of the island, due to the formula Equ.~(\ref{eq:islandformula}), and to the graviton mass as it indicates spontaneous diffeomorphism breaking.

More interestingly, in this scenario one is not able to easily measure the graviton mass using the asymptotic fall-off due to the shockwave. Though one is still able to measure it locally, as we discussed in Sec.~\ref{sec:meaningmass}. Hence, this suggests the irrelevance of the argument that graviton mass can only be measured asymptotically. This is another bonus of this scenario.
\\

\noindent\textbf{Projective Measurement:} Before we wrap the discussion in this section, we notice that it would be very interesting to understand what would happen if one performs a projective measurement on the bath. Such a projective measurement will significantly reduce, if not eliminate, the entanglement between the bath and the AdS. Therefore, both the entanglement island and the graviton mass will be corrupted if one uses the formula Equ.~(\ref{eq:islandformula}) and follows the calculation in Sec.~\ref{sec:massvsdiss}.  However, by definition, all the physics inside the island is encoded in the bath, so all this information must suddenly be transferred to somewhere as after the projective measurement the entanglement wedge of the CFT includes almost all, if not all, of the AdS spacetime. We leave a careful study of this question to future studies.

\subsection{Islands in Complicated Backgrounds and Complexity}\label{sec:complex}

Another interesting question regarding the graviton mass and entanglement island is that one might consider having an island in a complicated asymptotically AdS background which has enough features to define operators localized at the points inside the AdS to all orders in perturbation theory.  Thus, this might circumvent the consistency condition we considered in \cite{Geng:2021hlu} without the need for a graviton mass \cite{Folkestad:2023cze,Bahiru:2022oas}.\footnote{See also the bottom of page 16 in \cite{Antonini:2025sur}.}

In fact, this point has already been discussed to some extent in the original paper \cite{Geng:2021hlu}. The main points in \cite{Geng:2021hlu} were that these backgrounds are not relevant for large AdS black holes at late times, and for evaporating black holes one can consider the evaporation to be slow such that the operators dressed to the evaporation are highly delocalized in time. Thus, it is hard to imagine if these operators are of any physical relevance to the island. 

Motivated by this question, a sharper consistency condition of entanglement island is proposed in \cite{Geng:2025rov}. This has been discussed in Sec.~\ref{sec:necc} that operators in the island should excite the bath Hamiltonian and so be detectable with the bath Hamiltonian. The operators dressed to the background features inside the AdS certainly don't obey this consistency condition. Thus, this consideration completely resolves this issue of dressing to background features to avoid graviton mass. While we notice that a potential argument against this consistency condition is that in holography we usually consider highly chaotic systems, and for a chaotic system the energy spectrum becomes nonperturbatively dense above the ``black hole threshold". Thus, in a complicated background like a black hole with the bath dynamics also being chaotic, small excitations can be detected using the bath Hamiltonian only nonperturbatively due to the dense energy spectrum nearby. Therefore, one might conclude that the consistency condition proposed in \cite{Geng:2025rov} is not relevant and the operators in the island are highly complicated in the bath.\footnote{Some relevant discussions can be found in Section 2.5.1 of \cite{Engelhardt:2023xer}.} First, one should remember that, as we have emphasized in this paper, the island is really a property of the quantum state, not specific dynamics of the system, but chaos is a property of dynamics. Thus, this already suggests the resolution of this puzzle. More explicitly, the detailed nature of the bath is not relevant for the existence of the island and the nonvanishingness of the graviton mass. Thus, one can equally consider a bath with integrable dynamics for which there is no complicated observable and the energy spectrum is not dense anywhere. Therefore, with the relevant physics isolated, potential counter-arguments regarding the relevance of the consistency condition in \cite{Geng:2025rov} are cleanly addressed.

\subsection{Changing the Definition of Islands}
We note that attempts to provide counter-examples to massive islands exist, for example see Sec.3 of \cite{Antonini:2025sur}, by trying to change the definition of islands. In this subsection, we will review and analyze the two examples in Sec.3 of \cite{Antonini:2025sur} and remind the readers that the only existing concrete example in this spirit is still the one provided in \cite{Geng:2020fxl}. Thus, it is an interesting future direction to construct more examples.

First, let's remind ourselves of the definition of the island and articulate how it can be changed such that one could hope to have some nontrivial examples in massless gravity. Entanglement island has been defined carefully in \cite{Geng:2021hlu} as entanglement wedges with no intersection with the boundary of a gravitational universe and have a nontrivial component inside the gravitational universe. If this is the situation, the nontrivial component in the gravitational universe is called the entanglement island. The aim of this definition was to isolate the relevant physics in the black hole information paradox. A natural change of the above definition is to consider entanglement wedges which intersect the boundary of a gravitational universe and also have disconnected components inside the gravitational universe. To be precise, we will call the disconnected components in these cases \textit{pseudo islands}. See \cite{Geng:2025tba1} for a potential consistency condition of pseudo islands and how it can be satisfied using quantum wormhole operators.

Here are two proposed examples of pseudo islands in \cite{Antonini:2025sur}:

\section*{ACMP I}
\begin{figure}
    \centering
    \begin{tikzpicture} [scale=0.5,decoration=snake]
\draw[-,very thick,red] (0,5) arc (90:450:5);
 \draw[fill=black, draw=none, fill opacity = 0.5] (0,1) arc (90:450:1);
 \draw[-,very thick,green] (0,0.6) arc (90:450:0.6);
 \draw[-,very thick,green] (3.5355,3.5333) arc (90:270:1.8 and 3.5355);
 \draw[-,very thick,black] (-1.76775,1.76775) arc (160:200:5.168555);
 \draw[-,dashed,black] (-3.5355,3.5333) to (-1.76775,1.76775);
 \draw[-,dashed,black] (-3.5355,-3.5333) to (-1.76775,-1.76775);
 \draw[->,decorate,orange] (1.1,0) to (2.6,0); 
 \draw[->,decorate,orange] (0,1.1) to (0,2.6); 
 \draw[->,decorate,orange] (0,-1.1) to (0,-2.6); 
 \draw[->,decorate,orange] (0.777817,0.777817) to (1.83848,1.83848); 
  \draw[->,decorate,orange] (0.777817,-0.777817) to (1.83848,-1.83848); 
   \draw[->,decorate,orange] (-0.952628,0.55) to (-1.81865,1.05); 
   \draw[->,decorate,orange] (-1.81865,1.05) to (-0.7,1.4); 
   \draw[->,decorate,orange] (-0.952628,-0.55) to (-1.81865,-1.05); 
   \draw[->,decorate,orange] (-1.81865,-1.05) to (-0.7,-1.4); 
   \node at (6.2,0) {\textcolor{black}{$A$}};
    \end{tikzpicture}
    \caption{\small{This setup contains a small black hole in AdS with a mirror putting on the left of it. The mirror was designed to reflect the Hawking quantum. Due to this reflection, a subsystem like $A$ in the figure would get more Hawking radiation than if there is no mirror. Because of this enhanced capture of Hawking radiation, the QFT entanglement between the bulk subregion, bordered on $A$ and a candidate RT surface $\gamma_{A}$ (the big green curve), and the black hole, there might be another disconnected components of the RT surface (the small green circle) formed inside the black hole horizon following Equ.~(\ref{eq:islandformula}).}}\label{pic:example1}
    \end{figure}
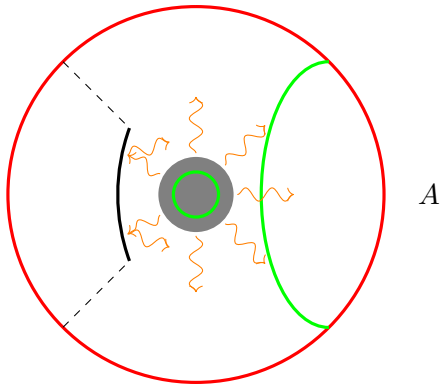

The first example considered in \cite{Antonini:2025sur} is depicted in Fig.~\ref{pic:example1}, where one has a small (evaporating) black hole in AdS together with a ``mirror" on its left. The purpose of the ``mirror" is to reflect the radiation from the black hole such that the boundary parts on the opposite side of the black hole to the mirror could be bombarded by an enhanced amount of Hawking radiation. Thus, if one considers a boundary subregion $A$ as depicted in Fig.~\ref{pic:example1} and takes a putative RT surface $\gamma_{A}$ as the big green curve, then the entanglement entropy of the bulk subregion in between $A$ and $\gamma_{A}$ will be enhanced by the mirror. This entanglement entropy is due to the entanglement between the Hawking radiation and the black hole. Thus, the hope is that, with this enhancement, this subregion entanglement entropy will be large enough such that a disconnected component of the entanglement wedge of $A$ will be developed inside the black hole as the result of Equ.~(\ref{eq:islandformula}). A putative disconnected component is drawn in Fig.~\ref{pic:example1} whose boundary is the small green circle. The authors in \cite{Antonini:2025sur} argued that this disconnected component must form if the boundary subregion $\mathcal{A}$ is large enough. This is a reasonable expectation and it would be very interesting to check explicitly if this is indeed the situation.

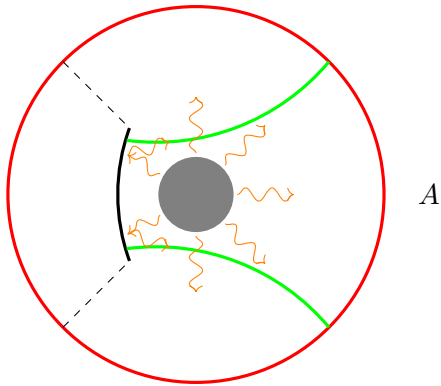
\begin{figure}
    \centering
    \begin{tikzpicture} [scale=0.5,decoration=snake]
\draw[-,very thick,red] (0,5) arc (90:450:5);
 \draw[fill=black, draw=none, fill opacity = 0.5] (0,1) arc (90:450:1);
 \draw[-,very thick,black] (-1.76775,1.76775) arc (160:200:5.168555);
 \draw[-,dashed,black] (-3.5355,3.5333) to (-1.76775,1.76775);
 \draw[-,dashed,black] (-3.5355,-3.5333) to (-1.76775,-1.76775);
 \draw[-,very thick,green] (3.5355,3.5333) arc (-40:-97.5:6);
  \draw[-,very thick,green] (3.5355,-3.5333) arc (40:97.5:6);
 \draw[->,decorate,orange] (1.1,0) to (2.6,0); 
 \draw[->,decorate,orange] (0,1.1) to (0,2.6); 
 \draw[->,decorate,orange] (0,-1.1) to (0,-2.6); 
 \draw[->,decorate,orange] (0.777817,0.777817) to (1.83848,1.83848); 
  \draw[->,decorate,orange] (0.777817,-0.777817) to (1.83848,-1.83848); 
   \draw[->,decorate,orange] (-0.952628,0.55) to (-1.81865,1.05); 
   \draw[->,decorate,orange] (-1.81865,1.05) to (-0.7,1.4); 
   \draw[->,decorate,orange] (-0.952628,-0.55) to (-1.81865,-1.05); 
   \draw[->,decorate,orange] (-1.81865,-1.05) to (-0.7,-1.4); 
   \node at (6.2,0) {\textcolor{black}{$A$}};
    \end{tikzpicture}
    \caption{\small{An equally probable, if not more likely, configuration of the entanglement wedge of $A$. The RT surface has two components ending on the ``mirror".}}\label{pic:example1cheaper}
    \end{figure}

However, here we would like to make a remark that the entanglement wedge could equally probably, if not more likely, be of a different form. This is because a much cheaper way to reduce the entanglement of the above subregion between $A$ and $\gamma_{A}$ is for that subregion to include the whole black hole, as is the dominant configuration of the minimization problem in Equ.~(\ref{eq:islandformula}) (see Fig.~\ref{pic:example1cheaper}). One such possibility is that the RT surface has two components both ending on the mirror. One might think that it is weird for the RT surface to end on this mirror. In fact, this is very conventional. One can remind oneself about the situation in the Karch-Randall braneworld, where the RT surface ends on the Karch-Randall brane because of the replica wormhole \cite{Geng:2024xpj} and in the conventional island story this gives entanglement islands \cite{Almheiri:2019hni,Geng:2020qvw,Chen:2020uac,Chen:2020hmv}. Moreover, from the bulk perspective, the Karch-Randall brane is a perfect mirror. Thus, one could already conclude from here that the RT surface of $A$ can end on the ``mirror". Here we notice that for the RT surface to end on the brane, it is not necessary for the brane to intersect the AdS boundary in the Lorentzian signature. For example, the configurations considered in \cite{Cooper:2018cmb,Antonini:2024bbm} have RT surfaces ending on an Karch-Randall brane residing in the black hole interior.\footnote{We thank Yikun Jiang for pointing out this example.} More precisely, in the replica derivation of the RT formula for $A$ there can exist a replica wormhole connecting the ``mirrors" on different copies resulting in the ``mirror" ending RT surface. In fact, it is not hard to see that if the boundary subregion $A$ is large enough, the entanglement wedge as depicted in Fig.~\ref{pic:example1cheaper} is preferred. Furthermore, as we discussed in the above paragraph, the argument in \cite{Antonini:2025sur} for the entanglement wedge in the form in Fig.~\ref{pic:example1} requires $A$ to be large enough. Hence, we can see that it is not so straightforward to conclude that we must have an entanglement wedge of the form in Fig.~\ref{pic:example1} in this setup considered in \cite{Antonini:2025sur}.

It would be very interesting to perform an explicit calculation to see if the entanglement wedge in the form as in Fig.~\ref{pic:example1} exists. We defer such a calculation to the future.

\section*{ACMP II}
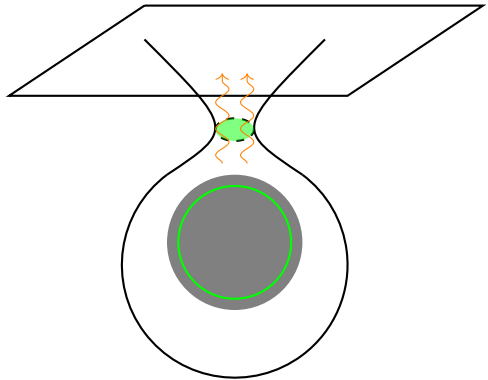
\begin{figure}
    \centering
    \begin{tikzpicture} [scale=0.3,decoration=snake]
\draw[-,thick, black] (-1,2) to (-7,-2) to (8,-2) to (14,2) to (-1,2); 
\draw[-,thick,black] (-1,0.5)..controls (3,-3.5).. (0,-5.5);
\draw[-,thick,black] (6,-5.5) ..controls (3,-3.5).. (7,0.5);
\draw[-,thick,black] (0,-5.5) arc (90+36.8698976 :90-36.8698976+360:5);
\draw[-,thick,dashed,black] (2.15,-3.5) arc (180:-180:0.85 and 0.5);
\draw[fill=black, draw=none, fill opacity = 0.5] (3,-5.5) arc (90:450:3);
\draw[-,green,thick] (3,-6) arc (90:450:2.5);
\draw[fill=green, draw=none, fill opacity = 0.5] (2.15,-3.5) arc (180:-180:0.85 and 0.5);
\draw[->,decorate,orange] (2.45,-5) to (2.45,-1);
\draw[->,decorate,orange] (3.55,-5) to (3.55,-1);
    \end{tikzpicture}
    \caption{\small{The dual bulk contains a small black hole evaporating inside the bag. One considers the boundary subregion as the part of the bag above the neck (the dashed curve). The black hole emits radiation going out of the bag region through the neck. Thus, following similar considerations as in the previous example, one might think that the entanglement wedge of this subsystem can contain a disconnected piece inside the black hole. In this expectation the RT surface has two components with one as the green surface bounded by the neck and the other the green sphere in side the black hole.}}\label{pic:example2bulk}
    \end{figure}

Another example is considered in \cite{Antonini:2025sur}, with the same spirit as the first example above. This example is also in the context of AdS/CFT. In this example, one considers the CFT to be living on a geometry that looks like a surface of a bag of gold. Then one considers the bulk to be an asymptotic AdS spacetime with the asymptotic boundary as this surface of the bag of gold and an evaporating black hole inside the bag part. The black hole is emitting radiation out of the bag, as depicted in Fig.~\ref{pic:example2bulk}. One considers the boundary subregion as the region above the neck. By the same consideration from \cite{Antonini:2025sur} as we reviewed in the previous example, one might think that at late times the entanglement wedge of this subregion could have a disconnected piece inside the black hole, as shown in Fig.~\ref{pic:example2bulk}.

However, the same consideration as we provided in the previous example indicates that the entanglement wedge can in fact equally probably, if not more likely, be of a different form, as depicted in Fig.~\ref{pic:example1cheaper}. In this case, the black hole is contained in the entanglement wedge of the boundary subsystem
we are considering.\footnote{This is also pointed out by Andreas Karch.} Thus, it is not so straightforward that the entanglement wedge will indeed have a disconnected component.

It would be very interesting to perform an explicit calculation to see if the entanglement wedge in the form as in Fig.~\ref{pic:example2bulk} exists. We defer such a calculation to the future.

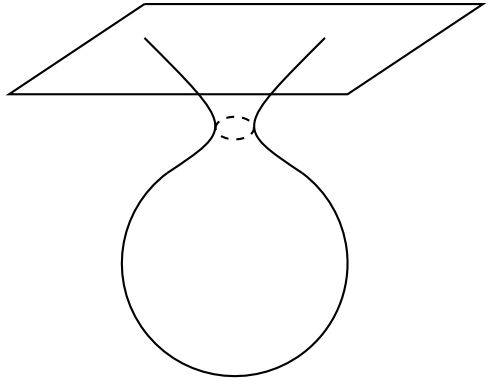
\begin{figure}
    \centering
    \begin{tikzpicture} [scale=0.3,decoration=snake]
\draw[-,thick, black] (-1,2) to (-7,-2) to (8,-2) to (14,2) to (-1,2); 
%\draw[-,thick,black] (-1,0.5)..controls (-2,-8.5) and (8,-8.5)..(7,0.5);
\draw[-,thick,black] (-1,0.5)..controls (3,-3.5).. (0,-5.5);
\draw[-,thick,black] (6,-5.5) ..controls (3,-3.5).. (7,0.5);
\draw[-,thick,black] (0,-5.5) arc (90+36.8698976 :90-36.8698976+360:5);
\draw[-,thick,dashed,black] (2.15,-3.5) arc (180:-180:0.85 and 0.5);
    \end{tikzpicture}
    \caption{\small{This setup considers the boundary CFT to be living on a background whose geometry is the surface of a bag of gold. One can think of it as a plane and with a disk (a ball in higher dimensions) on this plane pushed down.}}\label{pic:example2CFT}
    \end{figure}

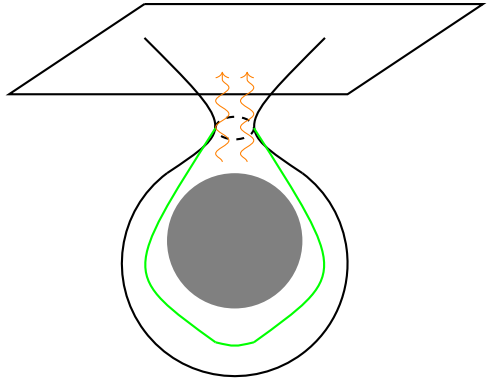
\begin{figure}
    \centering
    \begin{tikzpicture} [scale=0.3,decoration=snake]
\draw[-,thick, black] (-1,2) to (-7,-2) to (8,-2) to (14,2) to (-1,2); 
\draw[-,thick,black] (-1,0.5)..controls (3,-3.5).. (0,-5.5);
\draw[-,thick,black] (6,-5.5) ..controls (3,-3.5).. (7,0.5);
\draw[-,thick,black] (0,-5.5) arc (90+36.8698976 :90-36.8698976+360:5);
\draw[-,thick,dashed,black] (2.15,-3.5) arc (180:-180:0.85 and 0.5);
\draw[fill=black, draw=none, fill opacity = 0.5] (3,-5.5) arc (90:450:3);
\draw[-,thick,green] (2.15,-3.5)..controls (-2,-10)..(2.15,-13);
\draw[-,thick,green] (3.85,-3.5)..controls (8,-10)..(3.85,-13);
\draw[-,thick,green] (2.15,-13)..controls (3,-13.2)..(3.85,-13);
\draw[->,decorate,orange] (2.45,-5) to (2.45,-1);
\draw[->,decorate,orange] (3.55,-5) to (3.55,-1);
    \end{tikzpicture}
    \caption{\small{An equally probable, if not more likely, configuration of the entanglement wedge of the subregion is to just include the black hole. The green curve is the new RT surface.}}\label{pic:example2cheaper}
    \end{figure}

\section*{A Concrete Example}
With the recently proposed examples of pseudo islands analyzed in detail, we would like to remind the readers of a concrete example discovered in \cite{Geng:2020fxl,Geng:2023qwm}. 

\begin{figure}
\begin{centering}
\begin{tikzpicture}[scale=1]
\draw[-,very thick,black!100] (-2.5,0) to (0,0);
\draw[-,very thick,black!100] (0,0) to (2.5,0);
\draw[pattern=north west lines,pattern color=gray!200,draw=none] (0,0) to (-2.5,-1.875) to (-2.5,0) to (0,0);
\draw[pattern=north west lines,pattern color=gray!200,draw=none] (0,0) to (2.5,-1.675) to (2.5,0) to (0,0);
%violet!50
\draw[-,dashed,color=black!50] (0,0) to (-1.875,-2.5); 
\draw[-,dashed,color=black!50] (0,0) to (0,-2.875);
\draw[-,dashed,color=black!50] (0,0) to (1.875,-2.375); 
\draw[-,very thick,color=green!!50] (0,0) to (2.5,-1.6875); 
\draw[-,very thick,color=green!!50] (0,0) to (-2.5,-1.875);
%\draw[-] (-0.75,0) arc (180:217.5:0.75);
%\node at (-0.5,0.2) {$\mu$};
%\draw[-] (1.5,0) arc (0:-5.25:1.5);
\node at (0,0) {\textcolor{red}{$\bullet$}};
\node at (0,0) {\textcolor{black}{$\circ$}};
\draw[-,dashed,very thick, color=black!50] (-2,-1.5) arc (-140:-35:2.5);
\draw[-,dashed,color=black!50] (-1.5,-1.125) arc (-140:-35:1.875);
\draw[-,dashed,color=black!50] (-1,-0.75) arc (-140:-35:1.25);
\draw[-,thick,red] (0,0) to (0,-2.7);
\draw[-,thick,red] (0,0)..controls (-0.2,-1) and (-1.22,-1.3)..(-1.5,-1.125);
\draw[-,very thick,blue] (-1.5,-1.125) to (-2.5,-1.875);
\node at (-1.5,-1.125) {\textcolor{black}{$\bullet$}};
\node at (-1.2,-1.4) {\textcolor{black}{$\partial\mathcal{I}$}};
\end{tikzpicture}
\caption{\small{A constant time slice of an asymptotic AdS$_{d+1}$ with two Karch-Randall branes. The green surface denotes the brane and it has AdS$_{d}$ black hole geometry. The bulk geometry is a black string. The question we are interested in the entanglement entropy of an internal bipartition of the defect. There are three candidate RT surfaces for this entanglement entropy. The first candidate RT surface goes through the black hole interior connecting the two defects. The second candidate RT surface ends on the left brane and the third one ends on the right brane. For convenience we only draw the first two candidates. When a brane ending RT surface dominate, in the intermediate description there is an entanglement wedge which has two disconnected components. In this figure, such an entanglement wedge contains of whole right brane and the blue part on the left brane.}}
\label{pic:KRII}
\end{centering}
\end{figure}
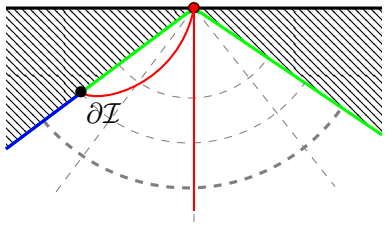

In this example, one considers a deformation of the original version of the Karch-Randall braneworld we reviewed in Sec.~\ref{sec:holographic}. This deformation introduces another Karch-Randall brane into the setup (see Fig.~\ref{pic:KRII}). One can think of this second Karch-Randall brane as a gravitational bath, and it cuts off the leftover bulk asymptotic boundary in the original Karch-Randall braneworld. If one remembers the discussion below Equ.~(\ref{eq:normgraviton}), one immediately realizes that with this second brane introduced, the zero mass eigenmode of the bulk graviton becomes normalizable. Thus, in this new setup, the holographic dual description of the (d+1)-dimensional bulk is that one has two gravitational AdS$_{d}$ spacetimes coupled to each other on their shared asymptotic boundary and the gravitational theory in this d-dimensional description is massless. One can think of the two AdS$_{d}$'s as the two Karch-Randall branes. Interestingly, here we can see that there is another holographic dual description, which is a (d-1)-dimensional conformal field theory on the defect. Interesting questions regarding holography can be studied in this setup.

The particular question we are interested in, regarding entanglement wedges that intersect the asymptotic boundary of a gravitational universe and contain disconnected components residing within a gravitational universe, is the entanglement entropy of a natural bipartition of the defect and the corresponding entanglement wedges in the d-dimensional description. To make the answer to this question go with a nontrivial Page curve, we will put the above setup in the black hole context. For this purpose, we consider a black string geometry in the (d+1)-dimensional bulk which is foliated by slices of d-dimensional AdS$_{d}$ eternal Schwarzschild black holes as depicted in Fig.~\ref{pic:KRII}. In this case, the d-dimensional defect description is the thermofield double state of a CFT$_{d-1}$. Then we can consider the entanglement entropy of an internal bipartition of this defect system. This entropy is computed by the candidate RT surfaces as in Fig.~\ref{pic:KRII}, for which the standard RT surface connects the two entangled defects on different asymptotic boundaries through the black string interior. Due to the growing of the interior volume of the black string, the entropy computed by this standard RT surface monotonically grows with time. This growth will be cut off at late times by the RT surface that connects the defects to one of the branes, and this RT surface has two disconnected components living on the two asymptotic exteriors of the bulk black string. This brane ending RT surface has been explicitly constructed in \cite{Geng:2020fxl} and we refer readers there for details. Hence, we would have a nontrivial time-dependent Page curve in this case, and at late times there is an entanglement wedge which contains one of the branes and another disconnected part living on the other brane as in Fig.~\ref{pic:KRII}. 

This is an explicit example of the type of entanglement wedges we discussed earlier in this subsection. We believe this type of entanglement wedges is very interesting, and we are looking for more explicit examples of it.

\section{Conclusions and Discussions}\label{sec:conclusion}

In this paper, we revisited the massive islands story \cite{Geng:2020qvw,Geng:2020fxl,Geng:2021hlu,Geng:2023zhq,Geng:2025rov} with interesting questions and potential confusions articulated and addressed. The key message we would like to convey is:
\begin{displayquote}
    \textit{\textbf{Graviton mass is not a bug but an interesting and essential feature of entanglement islands with a clear physical meaning.}}
\end{displayquote}

The connection between graviton mass and entanglement islands can be easily seen if one notices that they are both features of the quantum state as they are both due to quantum entanglement. For graviton mass, this can be seen by realizing that it is induced by the Higgs mechanism associated with the spontaneous diffeomorphism breaking. This spontaneous diffeomorphism breaking is indicated by the nontrivial local correlations between the gravitational AdS and the nongravitational bath. While, for the entanglement island, this is easily seen from the QES formula Equ.~(\ref{eq:islandformula}) which indicates that the formation of an island is due to the strong entanglement between the gravitational AdS and the nongravitational bath. However, we should emphasize that they are not tautological. This is because entanglement islands tell us how the information is encoded, and graviton mass explains why this encoding scheme is consistent and how it is realized. 

With the intuitive picture provided by the Karch-Randall braneworld \cite{Geng:2025rov}, we can see that the graviton mass is an explicit realization of the EPR=ER proposal \cite{Maldacena:2013xja}. This is because first the Goldstone vector boson associated with the spontaneous diffeomorphism breaking plays an essential role in understanding how the information about the island is consistently encoded in the nongravitational bath, and second in the Karch-Randall braneworld this Goldstone vector boson is dual to the extra-dimensional gravitational Wilson line that connects the island on the Karch-Randall brane to the nongravitational bath. Thus, we see that in a generic island model, without a semiclassical extra-dimensional dual, one should interpret the Goldstone vector boson as a quantum wormhole operator that connects the island and the bath through a quantum extra dimension. When the entanglement becomes strong enough such as in the Karch-Randall braneworld, this quantum extra dimension becomes semiclassical due to the condensation of the quantum wormholes \cite{Geng:2025rov}.

The graviton mass is a physically meaningful concept as it can be measured by a local observer inside the AdS without destroying the semiclassical geometric picture. Two such measurement protocols are provided in this paper. The first protocol is to measure the graviton propagator by studying the zeroth-order perturbative gravitational interaction between two matter sources.  There is a clean signature of the graviton mass, and one is able to measure the value of the graviton mass inside the AdS from this signature, as long as one is able to measure the value of the graviton mass near the AdS boundary. This addresses the potential confusion that ``one might only be able to measure the value of the graviton mass near the AdS boundary". The second protocol is a cheap protocol to show that the graviton mass is not zero inside the AdS. This is motivated by the fact that the graviton mass is generated by the Higgs mechanism. Thus, the nonvanishing graviton mass, i.e. the Higgs phase, can be shown by probing the energy scale of the spontaneous diffeomorphism breaking. A careful estimate of this scale following standard methods \cite{Arkani-Hamed:2002bjr} shows that this scale is much less than the Planck scale and much higher than the graviton mass in the physically relevant regime. Thus, it is a cheap deal to probe this scale, and this scale can be probed as in the standard Higgs mechanism by local scattering experiments of extra polarization modes of the gauge boson (gravity) with a sharp signature, as we discussed in Sec.~\ref{sec:scattering}. Moreover, we showed that the ``nonvanishing boundary flux" in the island model is a universal feature of the spontaneous symmetry breaking and Higgs mechanism, not a particular feature for the AdS Higgs mechanism.

After summarizing and analyzing various counter-arguments against the massive islands story at both the conceptual and technical levels, we realized that they can all be addressed by noticing the fact that the entanglement island is a property of the quantum state, as is the graviton mass.\footnote{More precisely, the nonzeroness of the graviton mass is a property of the quantum state as it is a reflection of spontaneous symmetry breaking and it requires only very mild assumptions of the dynamics, for example the symmetry is gauged. Though the exact value of the graviton might in some sense depends more heavily on the dynamics, for example it depends on the interaction strength.} Along the way, we also proposed many interesting questions for future studies. Moreover, we also suggested the search for a new type of entanglement wedges in holography, in the form as the one we found in \cite{Geng:2020fxl} and some recently proposed examples in \cite{Antonini:2025sur}.

\section*{Acknowledgments}
We thank Liam A. Fitzpatrick, \AA smund Folksted, Steve Giddings, Daniel Harlow, Tom Hartman, Ling-Yan Hung, Daniel Jafferis, Yikun Jiang, Andreas Karch, Emanuel Katz, Henry W. Lin, Juan Maldacena, Henry Maxfield, Rong-Xin Miao, Rashmish Mishra, Geoff Penington, Carlos Perez-Pardavila, Suvrat Raju, Lisa Randall, Martin Sasieta, Pushkal Shrivastava, Jia Tian, Jiuci Xu and Yoav Zigdon for helpful discussions. We thank Daniel Jaffeirs, Neeraj Tata and Pushkal Shrivastava for relevant collaborations. We thank the authors of \cite{Antonini:2025sur} (Stefano Antonini, Chang-Han Chen, Henry Maxfield and Geoff Penington) for sharing the prime version of their draft and Henry Maxfield and Geoff Penington for the correspondence about the prime version of \cite{Antonini:2025sur}. This paper comes out of illuminating questions we heard from many people in the last few years. The calculations in Sec.~\ref{sec:diffeobrokenscale} benefit from discussions with Lisa Randall. The question we discussed in Sec.~\ref{sec:massislocal} was originally pointed out by Juan Maldacena to us in 2023 and recently brought up by Yiming Chen. Similar questions are also asked by Martin Sasieta and Yoav Zigdon. The question at the beginning of Sec.~\ref{sec:notjustabox} about how graviton mass scales with the AdS length scale was originally pointed out by Daniel Harlow in a discussion. The question we discussed in Sec.~\ref{sec:complex} was pointed out to us by Daniel Harlow and Juan Maldacena in various discussions. We also thank Yikun Jiang, Henry Maxfield and Geoff Penington for comments on the draft. HG would like to thank the hospitality from the Aspen Center for Physics where the final stage of this work is performed. Research at the Aspen Center for Physics is supported by National Science Foundation grant PHY-2210452.  HG is supported by the Gravity, Spacetime, and Particle Physics (GRASP) Initiative from Harvard University and a grant from the Physics Department at Harvard University. 

\bibliographystyle{JHEP}

\bibliography{main}

\end{document}